\documentstyle[supertabular,epsf,rotate]{mn}
\newcommand{\mc}{\multicolumn}

\newcommand{\f}{\phantom{2}}
\newcommand{\ltsimeq}{\raisebox{-0.6ex}{$\,\stackrel 
        {\raisebox{-.2ex}{$\textstyle <$}}{\sim}\,$}} 
\newcommand{\gtsimeq}{\raisebox{-0.6ex}{$\,\stackrel 
        {\raisebox{-.2ex}{$\textstyle >$}}{\sim}\,$}} 
\input epsf.sty
\begin{document}
\title[The RLF of RLQs from the 7C Redshift Survey] 
{The radio luminosity function of radio-loud quasars from
the 7C Redshift Survey}

\author[Willott et al.]{Chris J.\ Willott, Steve Rawlings,  
Katherine M.\ Blundell and Mark Lacy\\ 
Astrophysics, Department of Physics, Keble Road, Oxford, OX1 3RH, U.K. \\
}
     
\maketitle

\begin{abstract}

We present a complete sample of 24 radio-loud quasars (RLQs) from the
new 7C Redshift Survey. Every quasar with a low-frequency (151 MHz)
radio flux-density $S_{151} > 0.5$ Jy in two regions of the sky
covering 0.013 sr is included; 23 of these have sufficient extended
flux to meet the selection criteria, 18 of these have steep radio
spectra (hereafter denoted as SSQs). The key advantage of this sample
over most samples of RLQs is the lack of an optical magnitude
limit. By combining the 7C and 3CRR samples, we have investigated the
properties of RLQs as a function of redshift $z$ and radio luminosity
$L_{151}$.

We derive the radio luminosity function (RLF) of RLQs and find that
the data are well fitted by a single power-law with slope $\alpha_{1}
= 1.9 \pm 0.1$ (for $H_{\circ}=50~{\rm km~s^{-1}Mpc^{-1}}$, $\Omega_
{\mathrm M}=1$, $\Omega_ {\Lambda}=0$). We find that there must be a
break in the RLQ RLF at ${\rm \log}_{10}(L_{151} /$ W Hz$^{-1}$
sr$^{-1}) \ltsimeq 27$, in order for the models to be consistent with
the 7C and 6C source counts. The $z$-dependence of the RLF follows a
one-tailed gaussian which peaks at $z=1.7 \pm 0.2$. We find no
evidence for a decline in the co-moving space density of RLQs at
higher redshifts.

A positive correlation between the radio and optical luminosities of
SSQs is observed, confirming a result of Serjeant et al. (1998). We
are able to rule out this correlation being due to selection effects
or biases in our combined sample. The radio-optical correlation and
best-fit model RLF enable us to estimate the distribution of optical
magnitudes of quasars in samples selected at low radio frequencies. We
conclude that for samples with $S_{151} \ltsimeq 1$ Jy one must use
optical data significantly deeper than the POSS-I limit ($R \approx
20$), in order to avoid severe incompleteness.

\end{abstract}

\begin{keywords}
Galaxies:$\>$active -- quasars:$\>$general -- quasars:$\>$emission lines 
\end{keywords}

\section{Introduction}

The variation of the space densities and properties of quasars with
redshift contains information about structure formation and evolution
at different epochs. Radio galaxies are important for similar reasons
and their relationship with radio-loud quasars (RLQs) is now under
scrutiny. According to orientation-based unified schemes, RLQs and
powerful radio galaxies are the same type of objects, their apparently
different properties being due to a systematic difference in the angle
between our line-of-sight and the radio jet axis (e.g. Scheuer 1987,
Antonucci 1993). In orientation-based unification, the broad line
region and quasar nucleus are hidden in radio galaxies behind an
optically-thick torus. If there is not a sharp cut-off in the optical
depth of the torus, then objects close to the transition angle between
radio galaxies and quasars may appear to be faint reddened quasars
(Baker 1997).

One of the major problems in trying to understand the collective
properties of a sample of objects is ensuring that the sample provides
reliable information on the parent population. Particularly in
astrophysics, care must be taken to minimise and understand selection
effects. For example, because of the steepness of the radio luminosity
function, there is a strong correlation between radio luminosity and
redshift in the 3CRR sample (Laing, Riley \& Longair 1983, hereafter
LRL). Hence, it is impossible to disentangle correlations with radio
luminosity from those with evolution with a single flux-limited
sample. Radio quasar samples selected at high-frequency contain mostly
flat-spectrum quasars (FSQs, $\alpha_{\mathrm rad} \leq 0.5$
\footnotemark), where $\alpha_{\mathrm rad}$ is evaluated at 1 GHz in
the source rest-frame. The radio emission from these objects is often
dominated by optically-thick radio cores, which are enhanced by
Doppler boosting because their jet axes lie close to the
line-of-sight. Therefore their total radio flux-densities (and hence
probability of being above a given flux-limit) are
orientation-dependent and often strongly time variable (e.g.
Seielstad, Pearson \& Readhead 1983). In contrast, the radio emission
from steep-spectrum quasars (SSQs, $\alpha_{\mathrm rad} > 0.5$) is
dominated by optically-thin extended emission and hence low-frequency
selected samples suffer far less greatly from these
orientation-dependent or variability biases. Some radio sources,
particularly compact-steep spectrum (CSS) sources (e.g. Fanti et
al. 1995) and Giga-Hertz Peaked Spectrum (GPS) sources (e.g. O'Dea,
Baum \& Stanghellini 1991), suffer synchrotron self-absorption at
low-frequencies and hence have radio spectra which flatten at
low-frequency, eg. 151 MHz. By evaluating the spectral indices at
rest-frame 1 GHz we avoid rejecting these source from our sample,
although the sensitivity of a survey to such sources is clearly a
function of both survey frequency and source redshift (see Blundell,
Rawlings \& Willott 1998a (Paper IV) for a fuller discussion).
 
\footnotetext{The convention for spectral index, $\alpha_{\mathrm
rad}$, is that $S_{\nu} \propto \nu^{-\alpha_{\mathrm rad}}$, where
$S_{\nu}$ is the flux density at frequency $\nu$.}

The luminosity function describes the co-moving density of a class of
objects as a function of redshift and luminosity. For quasars, one can
define the luminosity function in terms of the optical luminosity
function (OLF), X-ray luminosity function (XLF), or the radio
luminosity function (RLF). Most large quasar surveys have been
selected using optical, and more recently X-ray, criteria so the
quasar OLF and XLF are better constrained than the RLF. Boyle, Shanks
\& Peterson (1988) showed that the OLF can be fitted with a broken
power-law with the break luminosity evolving with redshift --- an
effect normally termed pure luminosity evolution (PLE). However,
Hewett, Foltz \& Chaffee (1993) and Miller et al. (1993) showed that
PLE was an inadequate model to explain the evolution of luminous
quasars. Hawkins \& V\'eron (1995) explained the break in the
power-law of Boyle et al. as due to incompleteness in their data.
Hewett et al. proposed that the power-law index of the OLF steepens as
the redshift increases. This is supported by the recent analysis of
the quasar OLF by Goldschmidt \& Miller (1998). The space density of
the most luminous quasars shows little or no evolution (Miller et
al. 1993), whilst the optically fainter quasars show significant
evolution (a factor of $\sim 50$ increase between $z \sim 0.5~{\rm
and}~ z \sim 2$). The XLF also appears to have a broken power-law form
(Maccacaro et al. 1991; Della Ceca et al. 1992). Boyle et al. (1994)
find the best-fitting model for the XLF is pure luminosity evolution
for $z<1.6$ and a constant co-moving density at higher redshift.
However, it must be remembered that these quasar samples selected in
the optical and X-ray wavebands may be prone to selection effects. If,
for example, many quasars are highly reddened (Webster et al. 1995),
then this would greatly alter the shape and evolution of the OLF, and
because of absorption by neutral hydrogen, the XLF.

One of the difficulties of determining the quasar RLF is that radio
surveys typically contain mostly radio galaxies, so all sources must
be reliably identified as quasar or radio galaxy. Dunlop \& Peacock
(1990) derived the RLF separately for flat- and steep-spectrum radio
sources. However, the high selection frequency (2.7 GHz) of the
samples used in this work meant a large fraction of flat-spectrum
sources in their study and hence the steep-spectrum RLF was more
poorly constrained. The steep-spectrum RLF fit of Dunlop \& Peacock
includes both radio galaxies and steep-spectrum quasars. No attempt
was made to separate these different types of object. Hence the
steep-spectrum {\em quasar} RLF has not previously been
well-constrained. Radio selected quasar samples usually have
additional optical selection criteria, which can be a cause of
incompleteness and selection biases. Since the 7C quasar sample has no
optical selection criteria it is complete, so in conjunction with the
3CRR sample of LRL it can be used to derive the RLF for the
low-frequency selected population which is dominated by SSQs.

There have been several suggestions of correlations between the radio
and optical properties of AGN. The close relationship between
narrow-line luminosity and low-frequency radio luminosity (e.g. Baum
\& Heckman 1989; Rawlings \& Saunders 1991; Willott et al. 1998b
(Paper III)) suggests a physical link between these two
properties. Correlations between the radio and optical luminosities of
RLQs have been found (e.g. Browne \& Murphy 1987) but incompleteness
worries mean that there are unknown selection effects which could
affect the results. Correlations between the radio and optical may be
due to orientation-dependent emission (e.g. Baker 1997). This is a
problem for flat-spectrum RLQ samples because often the radio
luminosities are dominated by a Doppler-boosted core and the optical
luminosities too may therefore be orientation-dependent (e.g. Wills \&
Lynds, 1978; Browne \& Wright 1985). Serjeant et al. (1998) presented
evidence for a radio-optical correlation for SSQ samples. They used
the 3CRR sample and a sample based on cross-matching the low-frequency
(408 MHz) MRC radio catalogue with APM data - the MAQS sample. This
sample has an optical magnitude limit as well as a radio flux
limit. They also calculated $B$-band absolute magnitudes using
observed $B$ magnitudes and $K$-corrections. For high-redshift
quasars, this radiation was emitted in the UV and so will be greatly
diminished if there is significant reddening of the quasar spectrum by
dust.
 
Excepting 3CRR, all other SSQ samples also have optical magnitude
limits, usually in the blue part of the spectrum (e.g. Vigotti et
al. 1997; Riley et al. 1998). Hence faint (and/or reddened) quasars
can be missed in these samples. If a substantial fraction of quasars
are reddened then quasar samples with bright magnitude limits will
miss them and an unrepresentative sample of the quasar population
(i.e. only the `naked' optically-bright ones) is seen. The only way to
be sure of avoiding these selection effects is to identify {\em all}
the radio sources in a flux-limited sample. However, this can be very
time consuming because most radio samples are dominated by radio
galaxies, not quasars.

Here we present a complete sample of 24 quasars from the new 7C
Redshift Survey. In this survey we have obtained optical spectra of
every radio source in two patches of sky (whose total area covers
0.013 sr), with a 151 MHz flux density $S_{151} > 0.5$ Jy. We made
high-resolution VLA maps of these sources to ensure reliable optical
identifications.  All 77 radio sources have optical / near-infrared
identifications and only 6 lack secure spectroscopic redshifts. All 6
of these are galaxies with estimated redshifts from multi-colour
photometry. Hence we can be confident that there is no significant
incompleteness in our sample of radio-loud quasars.

The layout of this paper (Paper II) is as follows. In Section 2 we
present the data on the complete 7C quasar sample. In Section 3 we
discuss our definition of a quasar in order to ensure that we are
comparing similar types of objects from the 3CRR and 7C samples. This
is necessary because the traditional definition of a quasar is no
longer tenable in the light of unified schemes. The RLQ radio
luminosity function is derived in Section 4 and compared with the
optically-selected quasar OLF. In Section 5 we discuss the evidence
for a correlation between the optical and radio luminosities of
steep-spectrum quasars. In Section 6 we use our knowledge of the
radio-optical correlation and the RLQ RLF to estimate the completeness
of quasar samples selected with low-frequency radio and optical flux
limits. The optical properties of the 7C quasar sample are to be
discussed elsewhere (Willott et al. in prep., Paper VIII). Also in this
forthcoming paper, we investigate the amount of reddening in the 7C
quasar sample and its impact on the completeness of quasar surveys.

\begin{table*}
\footnotesize
\begin{center}
\begin{tabular}{lcrcccc}
\hline\hline
\mc{1}{l}{Name} &\mc{1}{c}{Date}&\mc{1}{c}{Exposure}&\mc{1}{c}{Seeing}&\mc{1}{c}{Optical/NIR Position}&\mc{1}{c}{7C--optical}\\
\mc{1}{l}{ }      &\mc{1}{c}{}    &\mc{1}{c}{Time (s)}&\mc{1}{c}{(arcsec)}  &\mc{1}{c}{(B1950.0)}                  &\mc{1}{c}{separation (arcsec)}\\
\hline\hline	

 5C6.5      & 97Jan30  &  240 & 1.0 &  02 04 14.40~~  +33 22 20.1  & 3.4  \\
 5C6.8      & 97Jan30  &  240 & 1.0 &  02 04 39.63~~  +31 37 51.7  & 1.6  \\
 5C6.33     & 97Jan29  &  540 & 1.0 &  02 08 36.32~~  +33 34 18.5  & 2.8  \\
 5C6.34     & 96Sep04  &  540 & 1.0 &  02 08 38.50~~  +31 21 30.2  & 3.6  \\
 5C6.39     & 96Jan21  & 1080 & 1.1 &  02 08 53.76~~  +32 14 46.1  & 2.5  \\
 5C6.95     & 96Sep04  &  540 & 1.1 &  02 11 49.24~~  +29 36 40.0  & 8.3  \\
 5C6.160    & 97Jan29  &  540 & 0.9 &  02 14 44.80~~  +33 21 15.0  & 0.9  \\
 5C6.237    & 97Jan29  &  540 & 1.1 &  02 17 49.86~~  +32 27 23.3  & 2.0  \\
 5C6.251    & 97Jan29  &  540 & 1.1 &  02 18 50.44~~  +34 02 38.9  & 4.7  \\
 5C6.264    & 97Jan29  &  540 & 0.9 &  02 19 49.07~~  +33 43 41.3  & 2.2  \\
 5C6.282    & 96Sep04  &  540 & 1.0 &  02 21 40.78~~  +34 06 09.2  & 0.7  \\
 5C6.286    & 97Jan30  &  360 & 1.1 &  02 22 09.32~~  +31 45 49.2  & 4.4  \\
 5C6.287    & 96Sep04  &  540 & 0.9 &  02 22 19.95~~  +31 39 00.6  & 4.5  \\
 5C6.288    & 95Mar01  & 1080 & 1.1 &  02 22 19.33~~  +31 05 27.5  & 1.1  \\
 5C6.291    & 97Jan29  & 1080 & 1.1 &  02 23 09.74~~  +34 08 00.6  & 0.6  \\
 7C0808+2854& 97Jan30  &  540 & 1.2 &  08 08 32.11~~  +28 54 01.8  & 5.7  \\
 5C7.17     & 96Mar10  &  540 & 1.1 &  08 09 56.36~~  +27 00 57.5  & 6.2  \\
 5C7.70	    & 97Jan30  &  540 & 1.3 &  08 14 00.93~~  +29 27 30.9  & 1.3  \\
 5C7.85     & 96Mar10  &  540 & 1.2 &  08 14 45.58~~  +27 26 00.7  & 1.1  \\
 5C7.87     & 96Mar10  &  540 & 1.2 &  08 15 03.89~~  +24 58 27.3  & 6.0  \\
 5C7.95     & 95Mar01  & 1080 & 1.3 &  08 15 20.01~~  +24 45 05.4  & 6.8  \\
 5C7.118    & 96Mar10  &  540 & 1.2 &  08 16 15.06~~  +26 51 29.3  & 1.7  \\
 5C7.194    & 96Mar10  &  540 & 1.0 &  08 19 14.39~~  +25 48 09.8  & 0.9  \\
 5C7.195    & 95Mar01  & 2520 & 1.0 &  08 19 20.11~~  +25 15 37.8  & 0.8  \\
 5C7.230    & -        & -    & -   &  08 21 34.26~~  +24 48 28.5  & 3.0  \\
 7C0825+2930& 97Jan30  &  240 & 1.3 &  08 25 05.45~~  +29 30 17.9  & 3.2  \\
\hline\hline	     
\end{tabular}
\end{center}		  		
{\caption[Table of observations]{\label{tab:obktab} Log of $K$-band
observations of the 7C quasars and broad-line radio galaxies using
IRCAM3 on the UKIRT. The observations on 96Sep04 and 96Mar10 were part
of the UKIRT service observing program. All other observations were
made by us. The 7C--optical separation is the angular separation (in
arcsec) between the optical position and the radio position given in
the 7C catalogue. The position of 5C7.230 is from its identification
in the APM catalogue, since no $K$-band imaging of this object has
been obtained.}} \normalsize
\end{table*}

We assume throughout that $H_{\circ}=50~{\rm km~s^{-1}Mpc^{-1}}$. All
the calculations in this paper were done for three different cosmological
models to account for the current uncertainty in the values of the
matter density parameter ($\Omega_ {\mathrm M}$) and the cosmological
constant parameter ($\Omega_ {\Lambda}$). The models considered are I:
$\Omega_ {\mathrm M}=1$, $\Omega_ {\Lambda}=0$; II: $\Omega_ {\mathrm
M}=0$, $\Omega_ {\Lambda}=0$; III: $\Omega_ {\mathrm M}=0.1$, $\Omega_
{\Lambda}=0.9$.

\begin{table*}
\footnotesize
\begin{center}
\begin{tabular}{lccrcclc}
\hline\hline
\mc{1}{l}{Name} &\mc{1}{c}{Telescope +}&\mc{1}{c}{Date}&\mc{1}{c}{Exposure}&\mc{1}{c}{Slit width}&\mc{1}{c}{Slit PA}&\mc{1}{c}{Notes}&\mc{1}{c}{Previous}\\  
\mc{1}{l}{ }  &\mc{1}{c}{ Detector}&\mc{1}{c}{}& \mc{1}{c}{ time (s)} & \mc{1}{c}{ (arcsec)}& \mc{1}{c}{ ($^{\circ}$)}&\mc{1}{c}{}&\mc{1}{c}{reference}\\
\hline\hline	

 5C6.5      & WHT+ISIS  & 94Jan09 &  600 & 2.0 &  75 & variable cloud cover &  \\
 5C6.8      &      ''   & 94Jan08 &  200 & 2.0 & 255 & variable cloud cover &  \\
 5C6.33     &      ''   & 95Jan30 & 1800 & 2.0 &  60 &                      &  \\
 5C6.34     &      ''   & 94Jan09 &  500 & 2.0 &  75 & variable cloud cover &  \\
 5C6.39     &      ''   & 97Jan09 & 1200 & 2.5 & 141 &                      &  \\
 5C6.95     & WHT+FOS-2 & 91Jan15 &  475 & 4.0 &  66 &                      &  \\
 5C6.160    & WHT+ISIS  & 94Jan09 &  600 & 2.0 &  30 & variable cloud cover &  \\
 5C6.237    &      ''   & 94Jan08 &  300 & 2.0 & 255 & variable cloud cover & 1\\
 5C6.251    &      ''   & 94Jan08 &  300 & 3.0 & 255 & variable cloud cover &  \\
 5C6.264    &      ''   & 94Jan09 & 1200 & 2.0 &  30 & variable cloud cover &  \\
 5C6.282    &      ''   & 95Jan31 & 1300 & 2.0 &  81 &                      &  \\
 5C6.286    &      ''   & 94Jan09 &  600 & 2.0 &  30 & variable cloud cover &  \\
 5C6.287    &      ''   & 94Jan09 &  600 & 2.0 &  75 & variable cloud cover &  \\
 5C6.288    &      ''   & 95Jan31 & 1800 & 2.7 &  90 &                      &  \\
 5C6.291    &      ''   & 97Jan10 & 1800 & 1.0 &  24 &                      &  \\
 7C0808+2854&      ''   & 97Apr07 &  600 & 2.5 &  72 &                      & 2\\
 5C7.17     &      ''   & 95Jan29 &  900 & 3.1 & 165 &                      &  \\
 5C7.70	    &      ''   & 97Apr07 & 1200 & 1.5 & 120 &                      &  \\
 5C7.85     &      ''   & 94Jan10 &  730 & 2.0 & 280 & variable cloud cover &  \\
 5C7.87     &      ''   & 95Jan30 & 1700 & 3.1 &  60 &                      &  \\
 5C7.95     & LICK+KAST & 96Oct17 &  600 & 2.0 & 143 &                      &  \\
 5C7.118    & WHT+ISIS  & 97Apr07 &  300 & 2.5 & 137 &                      &  \\
 5C7.194    &      ''   & 97Apr07 &  300 & 2.5 & 172 &                      &  \\
 5C7.195    &      ''   & 95Feb01 & 1200 & 2.9 &  70 &                      &  \\
 5C7.230    &      ''   & 97Apr07 &  300 & 2.5 &  70 &                      &  \\
 7C0825+2930&      ''   & 97Apr06 &  900 & 2.5 & 170 &                      &  \\
\hline\hline
\end{tabular}
\end{center}		  		
{\caption[Table of observations]{\label{tab:obstab} Log of
spectroscopic observations of the quasars. The ISIS spectra cover the
range from about 3000 \AA~ to 8500 \AA~ with resolution approximately
10 \AA~. The FOS-2 spectrograph at the WHT was used for 5C6.95. The
spectrum of 5C7.95 was kindly obtained for us with the Kast
spectrograph on the 3-m Shane Telescope at Lick Observatory by Andrew
Bunker and Isobel Hook. Redshifts for 5C6.237 and 7C0808+2854 had
previously been determined in the literature . References: 1-Hook
et al. (1996), 2-Barthel, Tytler \& Thomson (1990). }} \normalsize
\end{table*}

\section{The 7C Redshift Survey}

Only brief details of the 7C Redshift Survey are given here; full
details will appear elsewhere (see also Rawlings et al. 1998; Willott
et al. 1998a; Blundell et al. 1998b, 1998c (Paper I). The 7C survey
was carried out with the Cambridge Low Frequency Synthesis Telescope
(CLFST) at 151 MHz (McGilchrist et al. 1990) at a resolution of 70
$\times$ 70 cosec(Dec) arcsec. The 7C Redshift Survey was pursued in
three separate patches of sky: 7C--I, 7C--II and 7C--III. In this
paper we consider the 7C--I and 7C--II regions, which were chosen by
Rossitter (1987) to overlap with the 5C6 and 5C7 fields of Pearson \&
Kus (1978), respectively.  The 7C--III region is centred on the North
Ecliptic Pole (overlapping with the 38 MHz selected sample of Lacy et
al. 1993) and has similar spectroscopic completeness to 7C--I and
7C--II. However the 7C--III objects lie above the declination limit of
the UKIRT and we have been unable to obtain $K$-band images of most of
these objects. For this reason the 7C--III region will not be
considered further in this paper. The 7C--I and 7C--II 151 MHz data
were subsequently re-analysed (and in the case of the 5C6 field
re-observed) by Julia Riley. VLA data, where available, were retrieved
from the archive (by kind permission of Barry Clark), or were recently
obtained by us: full details of all these radio observations and the
sample selection are described by in Paper I. The 7C--I region covers
an area of 0.0061 sr and contains 37 sources with $S_{151}>0.51$
Jy. The 7C--II region covers an area of 0.0069 sr and contains 40
sources with $S_{151}>0.48$ Jy. All the sources in the final sample
have now been mapped with the VLA to produce high-resolution radio
maps. These are necessary to resolve the lobes and core and get an
accurate position for the expected optical counterpart. The angular
separations between the 7C radio source positions and the optical
positions of the quasars are listed in Table 1. Note that these
separations are less than 10 arcsec in all cases (see Riley et
al. 1998 for further discussion of this point).

Despite the low selection frequency of the 7C survey (151 MHz), it is
possible that there could be Doppler-boosted quasars in the sample
whose extended flux alone would not exceed the sample flux limit. By
careful spectral fitting of the data, using integrated flux-densities
at a number of frequencies from many radio surveys and our own VLA
maps, we evaluated 1 GHz rest-frame spectral indices (hereafter
$\alpha_{\mathrm rad}$) for all the sources (as described in Blundell
et al. 1998a).  By consideration of these radio spectra and by
examining our own VLA maps (Paper I), we were able to identify any 7C
sources which exceeded the flux limit only because of the contribution
of their Doppler boosted cores.  Although six of the 24 quasars in the
7C sample have $\alpha_{\mathrm rad} \leq 0.5$, we believe that just
one (5C7.230) of these is promoted into the sample by such
orientation-dependent radio emission. The other five quasars in the 7C
sample with $\alpha_{\mathrm rad} \leq 0.5$ we believe belong to a
class of objects we term Core-JetS sources (CJSs); the details of
their radio structure and spectra are discussed in Paper I, but a
summary of their properties is as follows: these are quite compact
sources, whose spectra may be showing evidence of synchrotron
self-absorption, which when resolved with MERLIN or VLBI reveal
jet-type structures typically on both sides of the core.

We will use 23 7C RLQs in much of the subsequent analysis, comprising
18 which are SSQs (with $\alpha_{\mathrm rad} > 0.5$), and 5 which
have extended flux which exceeds the flux-limit but which have
$\alpha_{\mathrm rad} \approx 0.4$.

A similar analysis of the 3CRR sample (see Appendix B) leads to a sample of
RLQs consisting of 39 SSQs and one CJS (3C286). Possible reasons for the
differing SSQ:CJS ratios in the two samples will be explored elsewhere. Two
quasars from 3CRR (3C345 and 3C454.3) are excluded because they are promoted
into the sample by Doppler boosted core emission. 

\subsection{Source identification and $K$-band imaging}	

All 7C--I and 7C--II objects have been imaged at $K$-band
($2.2{\mathrm \mu m}$) for as long as required to obtain good
signal-to-noise on the radio source identification (except the
Doppler-boosted flat-spectrum quasar, 5C7.230). These images and their
analysis will be published elsewhere (Willott et al, in prep), but a
log of the observations is given in Table 1. Finding charts for the
entire 7C Redshift Survey were generated from the APM catalogue at the
RGO, Cambridge (Irwin, Maddox \& McMahon 1994). This catalogue
utilises scanned plates from the Palomar Observatory Sky Survey
(POSS). The astrometry of all images was fixed with the IRAF {\it
gasp} package using these finding charts. $K$ magnitudes were measured
with the IRAF {\it apphot}~ package using circular apertures of 5
arcsec diameter where possible. The infrared counterpart of the radio
source was observed in every single image. Hence, we are confident
that we have completely identified all members of the 7C Redshift
Survey. The measured quasar $K$-magnitudes are listed in Table 3.

\subsection{Optical spectroscopy}

About half of the 7C--I/7C--II quasars (13 out of 24) were identified
on the POSS-I plates as the likely optical counterpart of the radio
source, prior to spectroscopy. Some of the other quasars were
identified from prior imaging as candidate identifications. Where the
object was identified, an optical spectrum was taken at the position
of this object. If an object was not identified before time with a
spectrograph was available, then a spectrum was taken using blind
offsetting to the radio position, aligning the slit with the radio
axis (Rawlings, Eales \& Warren 1990). Most spectra were obtained with
the ISIS long-slit spectrograph at the WHT. Table 2 lists the details
of the quasar spectroscopy. The atmospheric conditions in January 1994
were not photometric since clouds affected the transparency.

The spectra were reduced in the IRAF package by subtracting the bias,
flat-fielding with a combination of twilight sky and tungsten lamp
spectra, wavelength calibration with comparison arc lamps, extinction
correction, background subtraction and flux-calibration using
spectrophotometric standards. One-dimensional spectra were extracted
from the 2-D frames using a FWHM aperture for a good signal-to-noise
ratio (snr) and a FWZI aperture for flux measurements. Atmospheric
absorption was corrected for by scaling the absorption from a standard
star spectrum by the airmass of the quasar observation. Distortions
where the red and blue spectra meet (due to the dichroic) were also
corrected using the flux-standard spectra. Final wavelength
calibration was achieved by applying a shift to the zero-point from
the 5577 \AA~ skyline.  Cosmic rays were identified in the 2-D spectra
and edited out of the 1-D spectra. Finally, the red and blue spectra
were joined together, averaging over 50 \AA~ where the snr of the
spectra were approximately equal.

\begin{table*}
\footnotesize
\begin{center}
\begin{tabular}{llccccccclclc}
\hline\hline
\mc{1}{l}{Name}& \mc{1}{c}{$S_{151}$}&\mc{1}{c}{$\log_{10}(L_{151}/$} &\mc{1}{l}{$\alpha_{\mathrm rad}$}&\mc{1}{c}{$z$} &\mc{1}{c}{$\alpha_{\mathrm opt}$}&\mc{1}{c}{$K$}&\mc{1}{c}{$b_{J}$}&\mc{1}{c}{$R$}&\mc{1}{c}{$M_{B}$}&\mc{1}{c}{$b_{J}$, $R$ phot.}\\
\mc{1}{l}{ }      & \mc{1}{c}{(Jy)}     &\mc{1}{c}{WHz$^{-1}$sr$^{-1}$)}  &\mc{1}{l}{ }   &    \mc{1}{l}{ } &\mc{1}{c}{ }  &\mc{1}{c}{}& \mc{1}{l}{ }              &\mc{1}{c}{}   &\mc{1}{c}{}       &\mc{1}{c}{source}\\
\hline\hline	

 5C6.5            & 0.767  & 26.432  & 0.96 & 1.038  &  0.1      &  $16.26 \pm 0.05$ & $19.15 \pm 0.25$ & $18.28 \pm 0.25$ & $-24.91$ & 1 , 1  \\
 5C6.8$^{\ddag}$  & 1.589  & 26.864  & 0.43 & 1.213  &  0.0      &  $16.28 \pm 0.06$ & $18.68 \pm 0.25$ & $17.83 \pm 0.25$ & $-25.52$ & 1 , 1  \\
 5C6.33           & 0.570  & 26.585  & 0.96 & 1.496  &  0.7      &  $18.23 \pm 0.19$ & $22.33 \pm 0.17$ & $21.08 \pm 0.05$ & $-23.20$ & 2 , 3  \\
 5C6.34           & 0.544  & 27.064  & 0.77 & 2.118  &  0.5      &  $15.94 \pm 0.04$ & $18.55 \pm 0.25$ & $17.94 \pm 0.25$ & $-26.65$ & 1 , 1  \\
 5C6.39           & 0.532  & 26.629  & 0.80 & 1.437  &  0.5      &  $17.87 \pm 0.08$ & $22.00 \pm 0.15$ & $21.35 \pm 0.07$ & $-23.09$ & 2 , 3  \\
 5C6.95           & 0.827  & 27.512  & 0.84 & 2.877  &  1.7      &  $16.04 \pm 0.04$ & $20.32 \pm 0.25$ & $19.25 \pm 0.25$ & $-26.88$ & 1 , 1  \\
 5C6.160          & 0.982  & 26.964  & 0.88 & 1.624  &  0.0      &  $17.95 \pm 0.14$ & $19.32 \pm 0.25$ & $18.82 \pm 0.40$ & $-24.72$ & 1 , 6  \\
 5C6.237$^{\ddag}$& 1.665  & 27.106  & 0.36 & 1.620  &  0.8      &  $15.76 \pm 0.03$ & $18.58 \pm 0.25$ & $17.24 \pm 0.25$ & $-26.58$ & 1 , 1 \\
 5C6.251          & 0.561  & 26.776  & 0.94 & 1.665  &  0.1      &  $17.66 \pm 0.11$ & $19.35 \pm 0.25$ & $18.93 \pm 0.25$ & $-24.84$ & 1 , 1  \\
 5C6.264          & 0.948  & 26.373  & 0.81 & 0.832&\hspace{0.15cm}1.3$^{*}$&$16.16\pm0.04$&$19.99\pm0.30$&$19.49\pm 0.04$ & $-23.50$ & 5 , 4  \\
 5C6.282          & 0.605  & 27.068  & 0.84 & 2.195  &  0.6      &  $18.20 \pm 0.18$ & $22.17 \pm 0.15$ & $21.70 \pm 0.14$ & $-23.91$ & 2 , 2  \\
 5C6.286          & 0.676  & 26.720  & 1.04 & 1.339  &  0.1      &  $17.62 \pm 0.20$ & $19.83 \pm 0.35$ & $19.20 \pm 0.35$ & $-24.34$ & 1 , 1  \\
 5C6.287          & 1.565  & 27.601  & 1.14 & 2.296  &  0.0      &  $16.11 \pm 0.04$ & $18.69 \pm 0.25$ & $17.81 \pm 0.14$ & $-26.74$ & 1 , 1  \\
 5C6.288$^{\ddag}$& 1.405  & 27.640  & 0.46 & 2.982&\hspace{0.15cm}2.9$^{*}$&$18.28\pm0.12$&$23.80\pm0.50$&$23.60\pm 0.50$ & $-24.42$ & 2 , 2  \\
 5C6.291$^{\ddag}$& 3.728  & 27.692  & 0.26 & 2.910  &  3.2      &  $16.06 \pm 0.03$ & $21.89 \pm 0.17$ & $20.54 \pm 0.12$ & $-26.62$ & 2 , 2  \\
 7C0808+2854      & 0.646  & 27.054  & 0.71 & 1.883  &  0.5      &  $16.06 \pm 0.04$ & $18.48 \pm 0.25$ & $17.86 \pm 0.25$ & $-26.39$ & 1 , 1  \\
 5C7.17$^{\dag}$  & 0.620  & 26.293  & 0.81 & 0.936  &  0.8      &  $17.37 \pm 0.07$ & $22.22 \pm 0.17$ & $21.59 \pm 0.14$ & $-21.88$ & 2 , 2  \\
 5C7.70	          & 2.084  & 27.809  & 0.97 & 2.617  &  3.1      &  $17.48 \pm 0.09$ & $22.81 \pm 0.18$ & $21.54 \pm 0.14$ & $-24.97$ & 2 , 2  \\
 5C7.85           & 1.420  & 26.716  & 0.82 & 0.995  &  1.8      &  $16.24 \pm 0.03$ & $20.68 \pm 0.23$ & $19.72 \pm 0.19$ & $-23.74$ & 2 , 2  \\
 5C7.87           & 0.843  & 27.132  & 0.86 & 1.764  &  0.1      &  $18.74 \pm 0.22$ & $21.47 \pm 0.16$ & $21.19 \pm 0.13$ & $-23.28$ & 2 , 2  \\
 5C7.95           & 0.672  & 26.671  & 0.89 & 1.203  &  0.5      &  $16.61 \pm 0.03$ & $20.07 \pm 0.25$ & $18.82 \pm 0.25$ & $-24.73$ & 1 , 1  \\
 5C7.118$^{\dag}$ & 1.417  & 26.174  & 0.69 & 0.527  &  $-0.1$~~ &  $15.75 \pm 0.03$ & $19.15 \pm 0.13$ & $19.06 \pm 0.11$ & $-22.77$ & 2 , 2  \\
 5C7.194          & 1.701  & 27.301  & 0.75 & 1.738  &  0.2      &  $15.91 \pm 0.03$ & $18.37 \pm 0.11$ & $18.10 \pm 0.10$ & $-26.20$ & 2 , 2  \\
 5C7.195          & 0.889  & 27.174  & 0.90 & 2.034  &  0.6      &  $17.71 \pm 0.05$ & $21.73 \pm 0.17$ & $21.27 \pm 0.15$ & $-24.16$ & 2 , 2  \\
 5C7.230$^{\ddag \ddag}$&0.634&26.538& 0.16 & 1.242  &  0.4      &          -        & $19.94 \pm 0.15$ & $19.35 \pm 0.13$ & $-24.04$ & 2 , 2  \\
 7C0825+2930$^{\ddag}$& 0.719& 26.989& 0.45 & 2.315  &  1.1      &  $17.55 \pm 0.15$ & $20.79 \pm 0.16$ & $20.63 \pm 0.14$ & $-24.85$ & 2 , 2  \\
\hline\hline	     
\end{tabular}
\end{center}		  		
{\caption[Table of observations]{\label{tab:phottab} Basic data and
photometry for the quasar sample. Radio luminosities and absolute
magnitudes were calculated assuming $\Omega_{M}=1$, $\Omega_
{\Lambda}=0$. $K$ magnitudes are all measured using 5 arcsec diameter
apertures except for 5C6.286, 5C6.288 \& 7C0825+2930 (3, 4 and 4
arcsec respectively, because of nearby objects). $b_{J}$ mags were
calculated from $B$ mags using $B\approx b_{J} - 0.14$ (e.g. Serjeant
et al. 1998). $\alpha_{\mathrm opt}$, $b_{J}$ and $R$ values corrected
for galactic reddening: 7C--I, $A_{B}=0.32$, $A_{R}=0.20$; 7C--II,
$A_{B}=0.25$, $A_{R}=0.16$.  $^{*}$The spectrum of 5C6.264 has a very
poor snr because of several magnitudes of extinction by cloud and that
of 5C6.288 only shows weak continuum because it is very
faint. Therefore their optical spectral indices have been calculated
assuming a power-law from their $R$ and $K$ magnitudes.
$^{\dag}$5C7.17 \& 5C7.118 are classified as broad-line radio galaxies
(BLRGs), not quasars, because they have $M_{B}>-23$ (see Section
3). $^{\ddag}$ These 5 quasars are core-jets sources (CJS; see Section
2). $^{\ddag \ddag}$5C7.230 is the only 7C quasar which does not have
sufficient {\em extended} flux at 151 MHz to be included as a member
of the complete sample. \\ Key to $b_{J}$, $R$ photometry:\\ 1 - APM
catalogue\\ 2 - measured from spectrum\\ 3 - WHT AUX port R band
image\\ 4 - McDonald Observatory, University of Texas at Austin,
$107^{\prime \prime}$ Telescope R band image\\ 5 - used measured $R$
mag. and assume $b_{J}-R=0.5$ adopting $\alpha_{\mathrm opt}=0.0$\\ 6
- not detected on the E plate, only on the O plate, assumed
$b_{J}-R=0.5$ to calculate $R$ magnitude.\\ }} \normalsize
\end{table*}

The final FWHM-aperture spectra are shown in Appendix A. A gaussian
fit was made to the lines to determine the line centres and
FWHMs. Where there was significant absorption present in the line
profile, this was removed before the fit. The redshift of each quasar
was determined from narrow line positions if possible, because these
are more accurately measured and believed to be at the systemic
redshift. Where narrow line centres could not be measured accurately,
a weighted mean of broad line centres was used to determine the quasar
redshift. For some objects blue-ward absorption of Ly$\alpha$
occurred, which shifts the measured line centre to higher
wavelengths. In these cases the redshift measured from Ly$\alpha$ was
not used in the redshift estimation procedure. Line fluxes were
measured with IRAF, by fitting a linear continuum and integrating the
flux above this continuum. The observed equivalent widths were also
measured in this way. The main source of error here is in accurately
determining the underlying continuum. Appendix A contains the emission
line data measured from the quasar spectra.

\subsection{Optical photometry}

$R$ and $b_{J}$ magnitudes have been measured for all the quasars and
are listed in Table 3. If possible, these values were taken directly
from the spectra. The FWZI aperture spectra were convolved with $R$
and $b_{J}$ filter transmission profiles. The flux within the filter
bandwidth was then measured. A similar procedure was followed with the
spectrophotometric standard, HD19445. This star has accurate
photometry in the $b_{J}$, $R$, $I$ system (Gullixson et al. 1995).
The measured quasar fluxes in each filter were then simply converted
to a magnitude.

For the quasars observed in non-photometric conditions, this approach
was not possible because a large fraction of the flux may be missing
in the spectra. The variable cloud cover on these nights caused up to
2 magnitudes of extinction in the spectra. However, the majority of
these quasars are detected on the APM $O$ and $E$ plates and have
measured magnitudes. In one case, the quasar was only detected on the
$O$ plate, so its $R$ magnitude was estimated from its optical
spectral index and $b_{J}$ magnitude. Two of the faint quasars were
imaged in $R$-band with the AUX port on the WHT, prior to
spectroscopy. These were 5C6.33 and 5C6.39 which were observed on 1995
Jan 30 and 31, respectively.  The exposure time for each image was
300s. A 300s $R$-band image of 5C6.264 was obtained at the
$107^{\prime \prime}$ telescope at the McDonald Observatory, operated
by the University of Texas at Austin, on 1995 Nov 20. These data were
reduced using standard procedures, as described in Lacy \& Rawlings
(1996).
 
\subsection{Galactic extinction corrections}
 
The 7C--I and 7C--II regions are at galactic latitudes, $b \approx
-30^{\circ}$ and $+30^{\circ}$, respectively. The amount of extinction
and its variability across the sample was investigated. This is
particularly important when one considers the intrinsic reddening of
quasars. IRAS 100${\mathrm \mu m}$ maps of the regions were
obtained from Skyview (http://skyview.gsfc.nasa.gov/) and the
positions of the 7C radio sources over-plotted.  The 100${\mathrm \mu
m}$ surface brightness (corrected for zodiacal emission) in any
direction is believed to be a very good indicator of the dust column
density and therefore the visual extinction. The two variables are
related by $A_{\mathrm V}=0.06~I_{100{\mathrm \mu m}}$, where
$I_{100{\mathrm \mu m}}$ is in units of MJy sr$^{-1}$ (Rowan-Robinson
et al. 1991).

The 100${\mathrm \mu m}$ maps show significant structure across each
region with several clumps of galactic cirrus with peak fluxes more
than twice the mean value. The mean extinction in each region is
$\langle A_{\mathrm V} \rangle=0.24$ (7C--I) and $\langle A_{\mathrm
V}\rangle =0.19$ (7C--II). None of the quasars lie in the direction of
the bright patches of cirrus. 5C6.95 is the quasar with the highest
visual extinction value, $A_{\mathrm V}=0.32$. The difference between
the $B$-band extinction to this quasar and the mean in 5C6 is 0.1
mag. This is much less than typical magnitude errors (such as those
from the APM), so we are justified in applying the same correction for
objects in the same field.

The effect of galactic extinction on the spectra was also
investigated. A flat spectrum ($\alpha_{\mathrm opt}=0$) was created
and then artificially reddened by various amounts (using the reddening
curve of Savage \& Mathis 1979). It was found that there is a
significant reddening of the initial flat spectrum. A power-law was
fitted to the reddened spectra in the region 4000-6000 \AA. For the
least reddened spectra (those in 7C--II), the reddened flat spectrum
was fitted with a power-law with spectral index, $\alpha_{\mathrm
opt}=0.22$. For 7C--I, $\alpha_{\mathrm opt}=0.27$ and the most heavily
galactic reddened quasar in the sample, 5C6.95, $\alpha_{\mathrm
opt}=0.35$. Hence, the scatter in $\alpha_{\mathrm opt}$ due to
variable galactic reddening is $< \pm 0.1$ and therefore not
significant compared with the errors in the determination of
$\alpha_{\mathrm opt}$. The spectra shown in Figure \ref{fig:spec}
have not been de-reddened.  However, the values of $b_{J}$, $R$ and
$\alpha_{\mathrm opt}$ in Table 3 have been corrected for galactic
extinction.

\section{Optical classification of radio-loud AGN}

Historically, quasars were defined as luminous objects with unresolved
`stellar' IDs and strong, broad (FWHM $> 2000$ km s$^{-1}$) emission
lines. In contrast, radio galaxies have a resolved optical appearance
and narrow (FWHM $\approx 1000$ km s$^{-1}$) or absent emission lines.
However, it is now apparent that this simple classification scheme is
inadequate. The increased sensitivity and resolution of optical and
near-IR imaging reveals host galaxies around quasars, particularly at
low $z$. Hence a weak quasar in a luminous galaxy will have a resolved
optical counterpart. Weak, broad emission lines have also been seen in
some radio galaxies. These objects are called broad-line radio
galaxies (BLRGs) and have absolute magnitudes, $M_{B}>-23$.
Spectropolarimetry of some radio galaxies has revealed broad lines
scattered into our line-of-sight (e.g. 3C234: Antonucci 1984, Cygnus
A: Ogle et al. 1997, 3C324: Cimatti et al. 1996), as expected in
orientation-based unified schemes. It is also clear that some quasars
have been reddened by dust so that the quasar is significantly dimmed
in the UV (e.g. 3C22: Rawlings et al. 1995, 3C41: Simpson, Rawlings \&
Lacy, in prep.). The identification of this type of reddened quasar
suffers from a redshift bias, because at higher redshifts optical
spectra sample shorter wavelengths where there is greater
extinction. Near-infrared data is required to identify reddened
quasars at high-$z$, e.g. the broad H-$\alpha$ line observed in the
near-IR spectrum of 3C22 (Rawlings et al. 1995). Therefore the simple
requirement that quasars have broad emission lines must be treated
with caution, because even with light reddening this becomes dependent
on redshift and whether NIR data is available.

Therefore we define quasars as objects with an unresolved nuclear
source which itself has $M_{B}<-23$ ($\Omega_ {M}=1$, $\Omega_
{\Lambda}=0$) after eliminating light from host galaxies and other
diffuse continuum (such as radio-aligned emission, e.g. McCarthy et
al. 1987). This value is the same as the limiting absolute magnitude
for quasars as defined in the quasar catalogue of V\'eron-Cetty \&
V\'eron (1996). $M_{B}<-23$ was decided upon after plotting $M_{B}$ vs
$L_{151}$ for all 3CRR and 7C Redshift Survey sources (see Section 5.1
for details of the $M_{B}$ calculation). There is a fairly clear
separation between narrow-line only objects, which are less luminous
than this, and objects with broad lines, which typically lie above
$M_{B}=-23$. BLRGs fall below $M_{B}=-23$ and are consequently not
classified as quasars. Further discussion of this classification
scheme is deferred to Paper VIII.

Two members of the 7C Redshift Survey are in the BLRG category: 5C7.17
and 5C7.118. They will not be considered here as members of the
complete quasar sample, although data is presented on them in Section
2. These objects are probably intrinsically weak AGN. They are not
significantly dust-reddened, optically-luminous quasars because their
optical-UV continua are blue (see Appendix A). The only remaining
problem with this definition is that heavily reddened quasars
($A_{\mathrm V} \gg 1$) will be excluded because their optical
luminosities are diminished to below our limiting absolute
magnitude. Such objects certainly exist (Simpson, Rawlings \& Lacy, in
prep.) and the consequences of this will be discussed in detail in
Paper VIII.

There are 2 sources in the 7C Redshift Survey which have optical
luminosities $M_{B}<-23$ and which are spatially compact at about
arcsec resolution, but for which we have no evidence of broad emission
lines. We believe that both these objects are quasars and they are
included as such in our sample. The spectrum of 5C6.264 was obtained
in poor atmospheric conditions and we believe this is the reason that
no broad lines are found in the spectrum. 5C6.288 is at high redshift
($z=2.982$) and shows only narrow lines in its optical spectrum. These
objects are briefly described in Appendix A and will be discussed in
more detail in Paper VIII. The optical spectrum of 5C7.195 also shows
strong narrow lines with only a marginal broad component to the C III]
emission line. This object is however known to be a lightly reddened
($A_{\mathrm V} \approx 1$) quasar (Willott et al. 1998a and Paper
VIII).

There are 8 objects traditionally classed as galaxies in the 3CRR
catalogue which have integrated absolute magnitudes, $M_{B}<-23$. All
of these are in the range $-24<M_{B}<-23$ and subtraction of the host
galaxy luminosity and/or extended emission provides a limit on their
unresolved optical component below our quasar definition value of
$M_{B}$. The most luminous galaxy 3C226 has been shown by the HST
imaging of Best, Longair \& R\"ottgering (1998) to contain a maximum
point-source contribution of 21\% (at the 90\% confidence
level). Therefore the maximum absolute magnitude of its nucleus is
$-22.0$, hence it is not by our definition a quasar.

As mentioned above, near-infrared observations of 3C22 and 3C41 (both
listed by Spinrad et al. 1985 as radio galaxies) reveal reddened
quasar nuclei. However, without de-reddening, these nuclei are fainter
than our limit. Unified schemes predict that all powerful radio
galaxies contain quasar nuclei which are heavily reddened. Therefore
we must exclude all reddened quasars with observed $M_{B}>-23$ nuclei
from our sample. 3C109 has often been quoted as a BLRG because at
$z=0.3056$, its host galaxy is visible in optical images. However, its
nucleus is brighter than our limit of $M_{B}=-23$, so we classify this
source as a quasar. Goodrich \& Cohen (1992) estimate that there are
$\simeq 3$ magnitudes of extinction towards the broad-line region of
3C109, so it is actually an intrinsically luminous quasar ($M_{B}
\approx -26$). It should also be noted here that some RLQs show
significant time variability in their nuclear optical luminosities and
this may lead to some uncertainty in object classification. The full
list of 3CRR quasars and BLRGs is in Appendix B.

\section{RLQ radio luminosity function}

\begin{figure*}
\epsfxsize=0.9\textwidth
\epsfbox{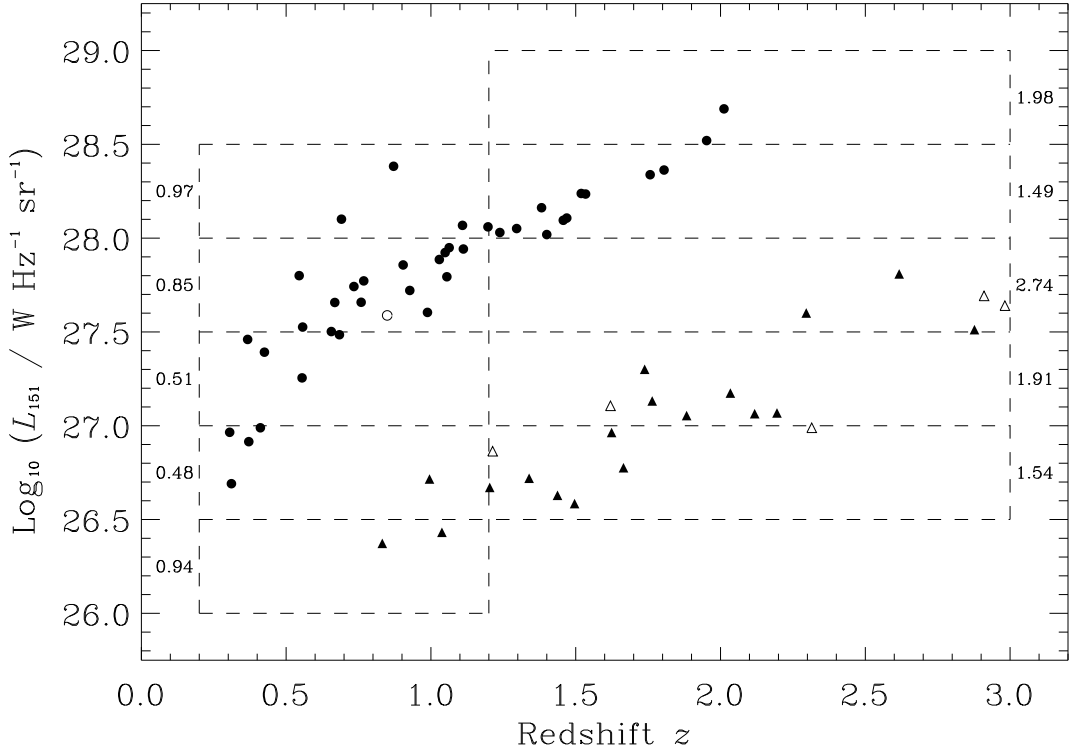}
{\caption[junk]{\label{fig:qsopz} The radio luminosity--redshift plane
for the 3CRR and 7C quasar samples. 3CRR quasars are plotted as
circles; filled are SSQs, open is the CJS. The triangles are 7C
quasars; filled are SSQs, open are CJSs. Note that the combination of
3CRR and 7C samples removes the strong luminosity--redshift
correlation present in a single flux-limited sample. The bins in
luminosity and redshift used in Section 4.1 are marked with dashed
lines. Next to each bin is the mean redshift of the quasars in the
bin. This plot is for cosmology I ($\Omega_ {M}=1$, $\Omega_
{\Lambda}=0$).}}
\end{figure*}

We can use complete radio-selected quasar samples to derive the radio
luminosity function (RLF) of radio-loud quasars. The dataset here
is a combination of the 40 quasars in the 3CRR sample (Appendix B) and
the 23 7C quasars with orientation-independent flux above the flux
limit (as discussed in Section 2). The 3 flat-spectrum quasars (3C345,
3C454.3 \& 5C7.230) which are promoted into the samples by their
prominent cores have not been included. Whilst this is only a small
sample ($N=63$), the simplicity of the selection criteria
compensate. The $S_{178} > 10.9$ Jy flux limit of the 3CRR sample was
recalculated at the same selection frequency as the 7C survey
(assuming low-frequency spectral index, $\alpha_{151}=0.8$), giving
$S_{151} > 12.43$ Jy. The radio luminosities of the 3CRR and 7C
sources were calculated at rest-frame 151 MHz using the spectral
fitting technique described in Blundell et al. (1998a). The radio
luminosity--redshift plane for both samples is shown in Figure
\ref{fig:qsopz}.

\subsection{Binned $1/V_{a}$ method}

The simplest way to estimate the luminosity function from a sample
with known fluxes and redshifts is to use the binned $1/V_{a}$ method
as described in Boyle et al. (1987). $V_{a}^{j}$ is the co-moving
volume available for a source $j$ in a bin with size $\Delta z$ (see
Avni \& Bahcall, 1980, for more details). $\rho(L,z)$ is the number of
sources per unit co-moving volume per unit $\log_{10}L_{151}$.

The RLF is given by
\begin{displaymath}
\rho(L,z) = \sum^{N}_{j=1} \frac{1}{V_{a}^{j}}~ \left( \Delta
\log_{10}L_{151} \right) ^{-1}.
\end{displaymath}
From Marshall (1985), the error bars on each bin are given by
\begin{displaymath}
\sigma = \left( \sum^{N}_{j=1} { \left( V_{a}^{j} \right) ^{-2}}
\right) ^{0.5} \left( \Delta \log_{10}L_{151} \right) ^{-1}.
\end{displaymath}
To reduce the effects of small number statistics, we chose to have two
redshift bins, equal in $\Delta \log_{10} (1+z)$. The redshift bins
are $0.2 \leq z \leq 1.2$ and $1.2 < z \leq 3.0$. Each bin in
luminosity is of size $\Delta \log_{10} L_{151}=0.5$. These bins are
shown in Fig.  \ref{fig:qsopz}.

\begin{figure}
\epsfxsize=0.52\textwidth
\begin{centering}
\epsfbox{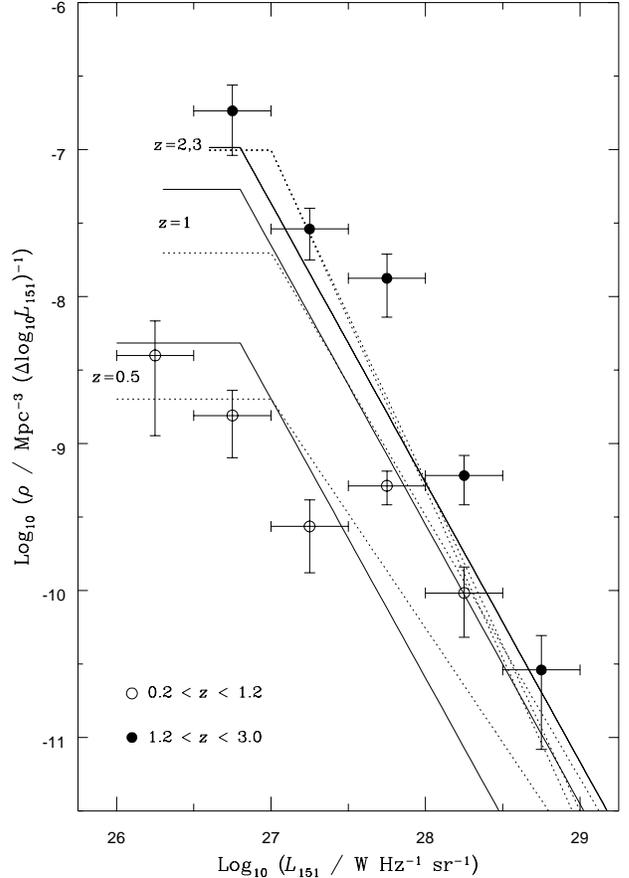}
\end{centering}
{\caption[junk]{\label{fig:qrlf} The radio luminosity function of RLQs
derived here (for cosmology I). The points with 1$\sigma$ error bars
are the binned RLF of Section 4.1. The solid lines are the
maximum-likelihood best-fit model C and the dotted lines model H,
plotted for several values of $z$.}}
\end{figure}

The results of the binned analysis are shown in Fig. \ref{fig:qrlf}
(for cosmology I). For each redshift bin, the RLF appears to be
consistent with a single power-law. The slope of the luminosity
function, $\alpha_{1}$, appears to steepen considerably from
$\alpha_{1}=0.8$ in the low-$z$ bin to $\alpha_{1}=1.8$ in the
high-$z$ bin. Marshall (1985) pointed out that with few redshift bins,
there will be significant evolution across each bin. Since our
coverage of the $L_{151}-z$ plane is non-uniform, this leads to the
mean redshift of objects in high-luminosity bins usually being greater
than that in the corresponding low-luminosity bins. Note there are
exceptions to this, e.g. the two highest luminosity, high-$z$ bins
contain only 3CRR quasars and consequently have lower mean redshifts
than the lower luminosity bins containing only 7C quasars. The peak in
mean redshift in the bin with the highest redshift 7C quasars is the
cause of the apparent bump in the binned high-$z$ RLF. The mean
redshifts of the sources in each bin are shown in
Fig. \ref{fig:qsopz}. Due to the strong evolution from $z \approx 0$
to $z \approx 2$ of RLQs, the main effect of this is an artificial
flattening of the power-law slopes in the binned method. Hence the
slopes quoted above are probably lower limits to the real values.

For cosmology II, the binned luminosity function does not appear to be
such a straight power-law as for cosmology I. The slope steepens
slightly at high luminosities. This is most noticeable for the
high-$z$ bins (but may be partly an artefact of the binning method as
discussed above). For cosmology III, the binned data has similar,
straight slopes as for cosmology I.

\subsection{Parametric model fitting}

The binned analysis used in Section 4.1 is not good statistical use of
complete sample data because information is lost when the objects are
binned. This is especially important when the most information
possible is to be extracted from a small dataset. An analytic model of
the luminosity function, $\rho(L,z)$, is required. The model should
have as few parameters as possible because of the small sample size
being used to constrain it. The first simple model tested (model A) is
a single power-law in luminosity with the redshift evolution specified
by a gaussian, i.e.
\begin{displaymath}
\rho(L,z) = \rho_{\circ} \left( \frac{L_{151}}{L_{\circ}} \right) ^{-\alpha_{1}}
\exp \left\{ - 
\frac
{\left( \frac{z-z_{\circ}}{z_{1}} \right)^{2} }
{2}
\right\},
\end{displaymath}
where $\rho_{\circ}$ is a normalising term, $L_{\circ}=10^{27}~{\rm
W~Hz^{-1}sr^{-1}}$, $\alpha_{1}$ is the power-law exponent,
$z_{\circ}$ is the redshift of the gaussian peak and $z_{1}$ is the
characteristic width of the gaussian. Thus there are only 4 free
parameters in model A. The use of a gaussian redshift distribution is
based on the observed redshift distribution of optically-selected
quasars (Shaver et al. 1998) and the global star-formation rate (Madau
et al. 1996).

\begin{figure}
\epsfxsize=0.52\textwidth
\begin{centering}
\epsfbox{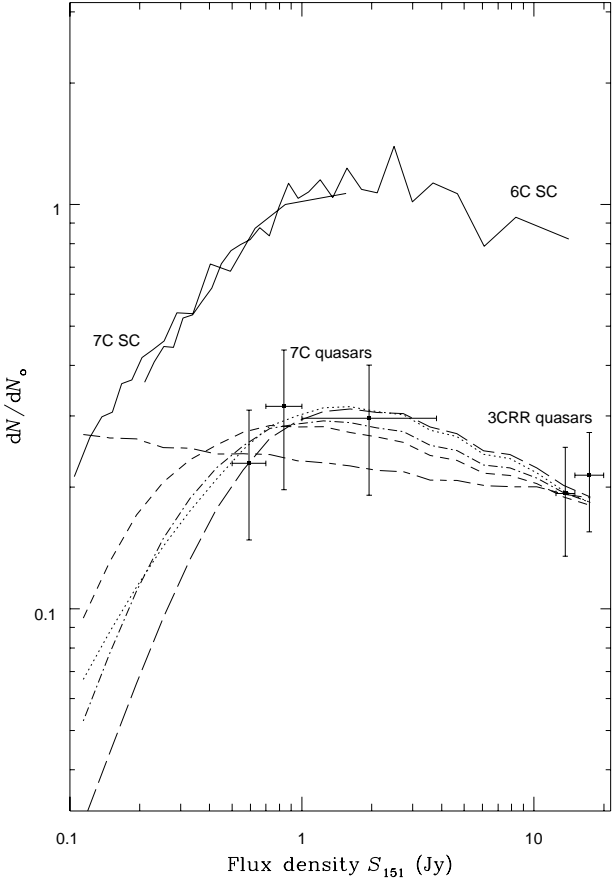}
\end{centering}
{\caption[junk]{\label{fig:source} The normalised differential source
counts from the 6C and 7C surveys are plotted here as solid lines and
labelled 6C SC and 7C SC. Note that these counts include all radio
sources, both quasars and radio galaxies. The solid circles with error
bars show the binned quasar source counts from the 7C and 3CRR quasar
samples described here. The source counts are normalised to the
differential source counts for a uniform distribution in a Euclidean
universe, such that $dN_{\circ}=2400 \left( S_{\mathrm
min}^{-1.5}-S_{\mathrm max}^{-1.5} \right)$, where $S_{\mathrm min}$
and $S_{\mathrm max}$ are the lower and upper flux limits of the
bin. The other lines are quasar source count predictions from RLF
models: A (short dash - long dash), B with $\log_{10} (L_{\rm break})
=26.6$ (short dash), B with $\log_{10} (L_{\rm break}) =26.8$ (dot
dash), B with $\log_{10} (L_{\rm break}) =27.0$ (long dash) and C with
$\log_{10} (L_{\rm break}) =26.8$ (dot).  }}
\end{figure}

To find the best-fit values of these parameters, the maximum
likelihood method of Marshall et al. (1983) was used. $S$ is defined
as $-2 \ln \mathcal{L}$, where $\mathcal{L}$ is the likelihood
function and terms independent of the model parameters can be
dropped. In this method, the aim is to minimise the value of $S$,
which is given by:
\begin{displaymath}
S = -2 \sum^{N}_{i=1} \ln [\rho(L_{i},z_{i})]+2 \int \hspace{-0.25cm}
\int \rho(L,z) \Omega(L,z) \frac{dV}{dz} dzdL
\end{displaymath}
where $(dV/dz)dz$ is the differential co-moving volume element,
$\Omega(L,z)$ is the sky area available from the samples for these
values of $L$ and $z$ and $\rho(L,z)$ is the model distribution being
tested. In the first term of this equation, the sum is over all the
$N$ sources in the combined sample. The second term is simply the
integrand of the model being tested and should give $\approx 2N$ for
good fits. The integration range for the second term is $0<z<5$, $24<
\log_{10} L_{151} <30$. Integrating the RLF up to $z=10$, instead of
$z=5$, in the maximum likelihood fitting made no difference to any of
the models, because the co-moving volume available becomes very small
at high redshift. Initial starting values of the free parameters were
estimated from the binned RLF. The parameters were optimised via the
downhill simplex method (Press et al. 1992). Errors on the best-fit
parameters were determined assuming a $\chi^{2}$ distribution of
$\Delta S$ ($\equiv S-S_{\mathrm min}$) (Lampton, Margon \& Bowyer
1976). The error on each parameter was calculated by setting $\Delta
S=1$ and allowing the other parameters to float. The errors quoted
here are 68\% confidence levels, as discussed in Boyle et
al. (1988). Table 4 contains the best-fit parameters of all the models
considered.

To assess the goodness-of-fit of a particular model the 2-dimensional
Kolmogorov-Smirnov (KS) test was applied. This multi-dimensional
version of the KS test was first proposed by Peacock (1983) and the
version used is that of Press et al. (1992). It gives the probability,
$P_{\rm KS}$, that the model distribution is a true representation of
the dataset. Note that the KS test can only be used to reject models,
not prove that they fit well. Indeed, for values of $P_{\rm KS}
\gtsimeq 0.2$ the probabilities may not be accurate and should not be
used to distinguish between models. Therefore, we also evaluate the
likelihood directly from the minimisation routine and use these to
compare well-fitting models. For model A, $P_{\rm KS} =0.03$ for
cosmology I and is even lower for cosmologies II \& III, which
indicates a poor fit. Another way to check the goodness-of-fit of a
RLF model is to use a Monte-Carlo simulation to create an artificial
$L-z$ plane. These artificial datasets can then be compared to the
actual $L-z$ plane for the 3CRR and 7C quasar samples. For model A,
the simulated data contained too many high radio luminosity sources
(i.e. in 3CRR) at both low ($z<0.3$) and high ($2<z<3$) redshift.

An additional constraint on our models down to $S_{151} \approx 0.1$
Jy comes from the 151 MHz source counts from the 7C and 6C surveys
(McGilchrist et al. 1990 and Hales, Baldwin \& Warner 1988,
respectively). Since these counts include both quasars and radio
galaxies, the source counts predicted from our quasar RLF models must
be significantly lower. Although the fraction of quasars in complete
samples at fluxes as low as 0.1 Jy is unknown, one would not expect it
to be vastly different from at 0.5 Jy. Models are considered
consistent with the source counts if the percentage of quasars
predicted at 0.1 Jy is in the range 20\%--40\% (the quasar percentage
in the 7C Redshift Survey ($S_{151} \ge 0.5$ Jy) is $\approx
30\%$). The single power-law model (A) predicts that the RLQ source
counts are greater than the total 7C source counts at 0.1 Jy. Figure
\ref{fig:source} plots the 6C and 7C total source counts and the
quasar source counts predicted by several RLF models. Taken together
with the poor goodness-of-fit test from the $L-z$ data, the source
counts constraint forces us to conclude that model A should be
rejected.

\begin{table*}
\footnotesize
\begin{center}
\begin{tabular}{ccccccccccllr}
\hline\hline 

\mc{1}{c}{Model} &\mc{1}{c}{Cos} &\mc{1}{c}{$N_{\mathrm par}$} &\mc{1}{c}{$\log_{10}$ ($\rho_ {\circ}$)}&\mc{1}{c}{$\alpha_{1}$}&\mc{1}{c}{$\alpha_{2}$}&\mc{1}{c}{$\log_{10}$($L_{\mathrm break}$)}&\mc{1}{c}{$z_{\circ}$}&\mc{1}{c}{$z_{1}$}&\mc{1}{c}{$k$}&\mc{1}{c}{$P_{\rm KS}$}&\mc{1}{c}{$\mathcal{L}$}&\mc{1}{r}{Q\%}\\

\hline\hline

A & I & 4 & $-7.199$ & $1.585$ & ---     &  ---   & $2.01$ & $0.505$ & ---    & $0.03$   & $10^{-7}$ & $>100$ \\
B & I & 4 & $-6.944$ & $1.875$ & ---     & $26.8$ & $2.16$ & $0.587$ & ---    & $0.44$   & $0.3$     & $25$ \\
C & I & 4 & $-7.150$ & $1.900$ & ---     & $26.8$ & $1.66$ & $0.436$ & ---    & $0.49$   & $1$       & $35$ \\
D & I & 3 & $-9.621$ & $1.338$ & ---     &  ---   & ---    & ---     & $3.56$ & $10^{-9}$& $10^{-25}$& $40$ \\
E & I & 3 & $-9.276$ & $2.105$ & ---     & $27.2$ & ---    & ---     & $3.97$ & $0.006$  & $10^{-10}$& $40$ \\
F & I & 4 & $-10.133$& $1.885$ & ---     & $26.8$ & $1.31$ & ---     & $8.10$ & $0.65$   & $0.2$     & $30$ \\
G & I & 5 & $-6.629$ & $1.334$ & $0.416$ & $27.0$ & $2.49$ & $0.647$ & ---    & $0.69$   & $6$       & $25$ \\
H & I & 5 & $-6.787$ & $1.329$ & $0.459$ & $27.0$ & $2.13$ & $0.548$ & ---    & $0.85$   & $10$      & $30$ \\
A & II& 4 & $-7.225$ & $1.576$ & ---     &  ---   & $2.02$ & $0.511$ & ---    & $10^{-4}$& $10^{-8}$ & $>100$ \\
B & II& 4 & $-6.730$ & $1.956$ & ---     & $27.2$ & $2.28$ & $0.629$ & ---    & $0.66$   & $0.6$     & $30$ \\
C & II& 4 & $-6.897$ & $1.976$ & ---     & $27.2$ & $1.84$ & $0.495$ & ---    & $0.58$   & $1$       & $35$ \\
D & II& 3 & $-9.553$ & $1.278$ & ---     &  ---   & ---    & ---     & $3.00$ & $10^{-8}$& $10^{-24}$& $30$ \\
E & II& 3 & $-8.416$ & $2.308$ & ---     & $27.6$ & ---    & ---     & $3.96$ & $0.03$   & $10^{-7}$ & $70$ \\
F & II& 4 & $-9.756$ & $1.961$ & ---     & $27.2$ & $1.45$ & ---     & $7.23$ & $0.75$   & $0.5$     & $35$ \\
G & II& 5 & $-6.302$ & $1.657$ & $0.261$ & $27.4$ & $2.69$ & $0.699$ & ---    & $0.44$   & $0.8$     & $30$ \\
H & II& 5 & $-6.525$ & $1.761$ & $0.228$ & $27.4$ & $2.16$ & $0.565$ & ---    & $0.47$   & $1.1$     & $35$ \\
A &III& 4 & $-7.334$ & $1.558$ & ---     &  ---   & $2.06$ & $0.508$ & ---    & $10^{-5}$& $10^{-8}$ & $>100$ \\
B &III& 4 & $-6.786$ & $1.934$ & ---     & $27.3$ & $2.34$ & $0.631$ & ---    & $0.65$   & $0.8$     & $30$ \\
C &III& 4 & $-6.936$ & $1.946$ & ---     & $27.3$ & $1.99$ & $0.531$ & ---    & $0.84$   & $1$       & $35$ \\
D &III& 3 & $-9.941$ & $1.291$ & ---     &  ---   & ---    & ---     & $3.74$ &$10^{-12}$& $10^{-22}$& $30$ \\
E &III& 3 & $-8.704$ & $2.212$ & ---     & $27.7$ & ---    & ---     & $4.31$ & $0.05$   & $10^{-6}$ & $60$ \\
F &III& 4 & $-9.963$ & $1.928$ & ---     & $27.3$ & $1.50$ & ---     & $7.37$ & $0.78$   & $0.4$     & $30$ \\
G &III& 5 & $-6.262$ & $1.590$ & $0.285$ & $27.5$ & $2.83$ & $0.703$ & ---    & $0.51$   & $1.1$     & $30$ \\
H &III& 5 & $-6.493$ & $1.676$ & $0.251$ & $27.5$ & $2.39$ & $0.602$ & ---    & $0.54$   & $1.1$     & $30$ \\         
\hline\hline         
\end{tabular}

Errors for model C (cosmology I):~ $\log_{10}$ ($\rho_ {\circ}$);
$-0.12, +0.13$,~ $\alpha_{1}; -0.09, +0.09$,~ $z_{\circ}; -0.19, +0.22$,~ $z_{1}; -0.06, +0.07$.
\end{center}              
{\caption[Table of observations]{\label{tab:qrlftab}Best-fit
parameters for model RLQ RLFs, as described in Section 4.2. Cos is the
cosmological model assumed, $N_{\mathrm par}$ is the number of free
parameters in the fit, $\mathcal{L}$ is the likelihood from the fit,
relative to model C for each cosmology and Q\% is the quasar
percentage predicted at 0.1 Jy from the source counts. A brief
description of each model is given below (see Section 4.2 for more
details).\\
A: single power-law RLF, 2-tailed gaussian $z$ dependence.\\ 
B: single power-law RLF with low luminosity break, 2-tailed gaussian $z$ dependence.\\ 
C: single power-law RLF with low luminosity break, 1-tailed gaussian $z$ dependence.\\ 
D: single power-law RLF, $(1+z)^{k}$ $z$-dependence. \\ 
E: single power-law RLF with low luminosity break, $(1+z)^{k}$ $z$ dependence.\\ 
F: single power-law RLF with low luminosity break, $(1+z)^{k}$ $z$ dependence with $z_{\mathrm max}=z_{o}$.\\ 
G: $z$ dependent power-law RLF with low luminosity break, 2-tailed gaussian $z$ dependence.\\ 
H: $z$ dependent power-law RLF with low luminosity break, 1-tailed gaussian $z$ dependence.\\ }}
\normalsize
\end{table*}

The steep-spectrum RLF of Dunlop \& Peacock (1990) (which includes
radio galaxies and quasars) could be fitted well by a broken power-law
with the break at $\log_{10} L_{2.7} \approx 25.5$. This is equivalent
to $\log_{10} L_{151} \approx 26.5$ (assuming $\alpha_ {\mathrm rad}=
0.8$). The over-prediction of the RLQ source counts by model A could
be caused by the lack of a break in the luminosity function. This is
because samples selected with $S_{151} \geq 0.1$ Jy, would contain low
luminosity sources ($\log_{10}L_{151} \approx 26$) out to redshifts,
$z \approx 1.5$. Therefore, the excess counts are due to the increased
number density of low radio luminosity sources in models with an
unbroken power-law.

Therefore, our next model (B) is a broken power-law in luminosity with
a break at $L_{\rm break}$. Unfortunately, we are unable to derive the
value of $L_{\rm break}$ or the power-law index below the break with
our complete sample data, because we have so few quasars at these low
radio luminosities. We assume that the RLF is flat ($\alpha_{1}=0$)
below the break. The value of $L_{\rm break}$ can be estimated using
the 6C/7C source counts, because the counts at 0.1 Jy are very
sensitive to it. For cosmology I we can constrain $L_{\rm break}$ for
this model to $26.7 < \log_{10} (L_{\rm break}) < 26.9$ (see Fig.
\ref{fig:source}). The quasar percentages for ${\log}_{10} (L_{\rm
break}) =26.6$ and ${\log}_{10} (L_{\rm break}) =27.0$ are 45\% and
15\%, respectively. ${\rm Log}_{10} (L_{\rm break}) =26.8$ was assumed
since it gives a quasar fraction at 0.1 Jy similar to that in the 7C
sample considered here. Similarly ${\rm log}_{10} (L_{\rm break})
=27.2$ and $27.3$ for cosmologies II and III, respectively.  Note that
the value of ${\log}_{10} (L_{\rm break})$ is not a free parameter in
our fits - it is constrained by the source counts, not the $L_{151}-z$
data. Maximum likelihood fits to the 4 free parameters of model B gave
values of $P_{KS}$ in the range 0.4--0.7 for all cosmologies which
indicates that the model fits our data fairly well.

\begin{figure*}
\epsfxsize=0.9\textwidth
\begin{centering}
\hspace{0.7cm}
\epsfbox{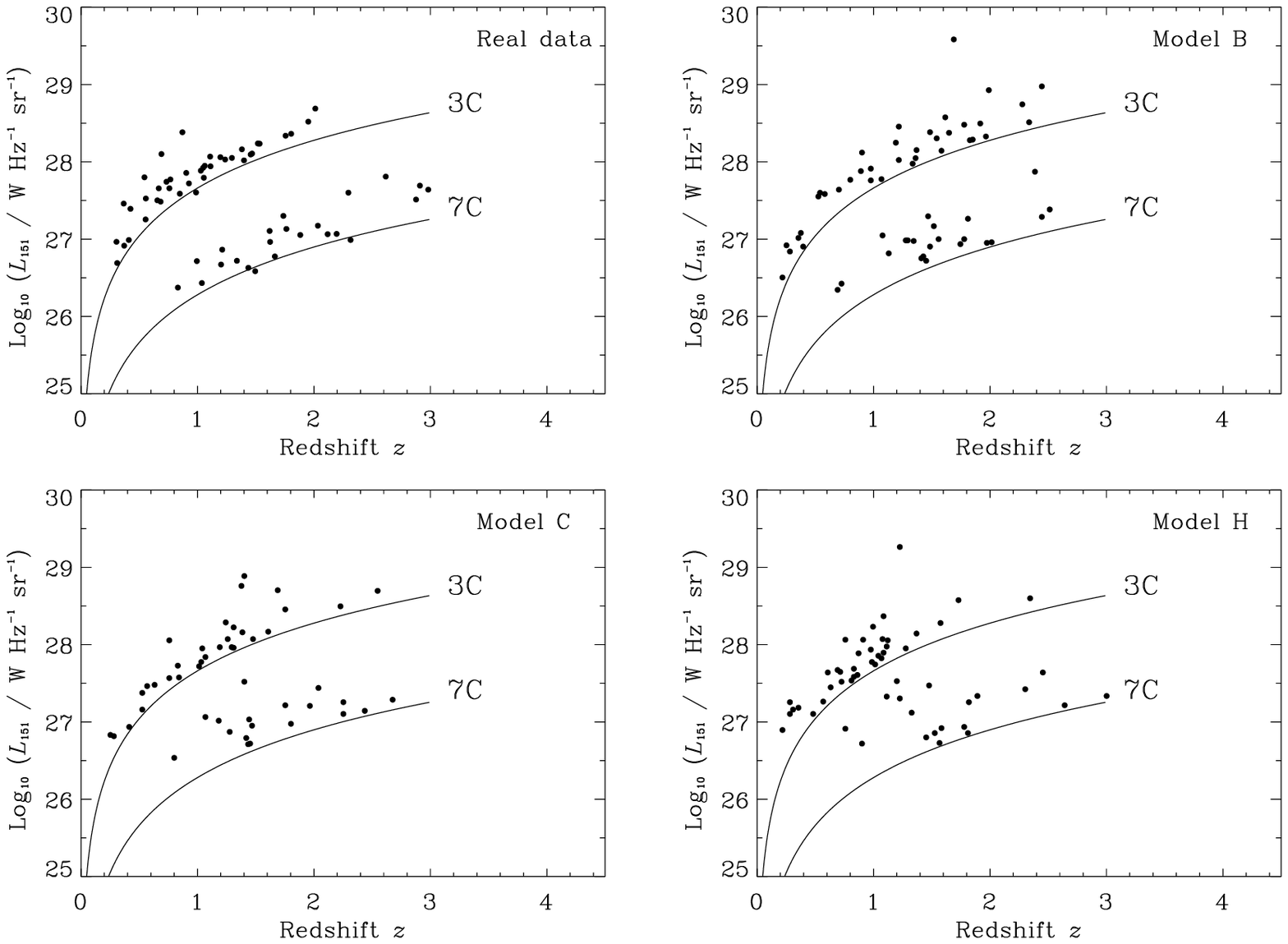}
\end{centering}
{\caption[junk]{\label{fig:montepz} The plot at top-left shows the
$L_{151}-z$ plane for the 3CRR and 7C quasar samples. The other 3
plots show Monte-Carlo simulations of the $L_{151}-z$ plane for
samples with the same flux limits and sky areas as the 3CRR and 7C
samples. These simulations were generated using the best-fit RLF
models B, C and H. All plots are for cosmology I.
}}
\end{figure*}

The functional form of the redshift evolution of the RLF in these
models is symmetric in $z$ about $z_{\circ}$. This form was not based
on theoretical predictions, but merely because it appears to be a good
fit to the increase in density from $z=0$ to $z \approx 2$ (although
clearly $\rho$ must decline at some high value of $z$). In the fit
using model B above, $z_{\circ}=2.2$ (for cosmology I). There are only
6 quasars in the 3CRR/7C sample with a redshift greater than this.
Therefore, we cannot claim that there is a symmetric decline in the
space density at $z>z_{\circ}$, just because we get a high probability
from our goodness-of-fit test. To test whether a high redshift decline
in $\rho$ is necessary, we repeated the model-fitting procedure using
a one-tailed gaussian in $z$, such that $\rho (z)$ does not decline at all
in the region $z_{\circ} \leq z \leq 5$ (and $\rho=0$ beyond $z=5$, as
before). This new model (C) fits the data equally well with similar KS
probabilities and a slightly greater likelihood. The peak redshift
for C ranges from 1.7 for cosmology I to 2.0 for cosmology III. Hence
we have no evidence for a decline in $\rho$ at $z\sim 2$ and we adopt
model C as a slightly better fit to the data than B.

Many derivations of the OLF and XLF have used the functional form
$(1+z)^{k}$ to describe the redshift evolution of the quasar
population (e.g. Boyle et al. 1994; Goldschmidt \& Miller 1998). We
have replaced the gaussian redshift term from model A with this new
term to make model D. Note that this reduces the number of free
parameters in our fit from 4 to 3. Model D provides a very poor
fit. The KS probabilities are lower than $10^{-8}$ for all
cosmologies. Monte-Carlo simulations of this model create many sources
at low redshift and high luminosities, completely unlike the actual
$L_{151}-z$ distribution.  As before, we now include a low-luminosity
break in the RLF (model E, analogous to model B). This model also fits
poorly with $P_{\mathrm KS}<0.05$ for all cosmologies. The relative
likelihoods between model E and model C are $<10^{-6}$. The
low-luminosity break in model E is required to be 0.4 dex greater than
models B \& C and even then the percentage of quasars in the source
counts at 0.1 Jy are $> 50\%$. Hence model E must be rejected. The
reason for its poor fit is that the number density of quasars continues
increasing up to high redshifts. This is illustrated by the
Monte-Carlo simulations which show many sources at $z>3$, unlike the
real data.

Boyle et al. (1994) find a good fit to the XLF with $(1+z)^{k}$
evolution if one allows a maximum redshift for this evolution,
$z_{\circ}$, such that $\rho$ is constant at $z>z_{\circ}$. This
should solve the problem of model E -- too many high redshift
quasars. Model F implements this form for the evolution and
consequently there are now 4 free parameters in the fit. Model F is
now extremely similar to model C, in that in both $\rho$ increases up
to $z_{\circ}$ and then remains constant. The fits for model F do
indeed have similar parameter values to model C. The KS probabilities
are in the range $0.6-0.8$. Model F agrees well with the source counts
(with ${\log}_{10} (L_{\rm break}) =26.8$, as for model C) and
Monte-Carlo simulations. The relative likelihoods of model F to model
C are a factor of 2 to 5 lower, so we retain model C as the
best-fitting model.

Our binned RLQ RLF has a power-law in $L_{151}$ which apparently
steepens as the redshift increases. Therefore, the next logical step
is to include this possibility into our RLF (model G). The power-law
part of the model RLF is now
\begin{displaymath}
\left( \frac{L_{151}}{L_{\circ}} \right) ^{-~ \left( \alpha_{1} +
\alpha_{2}z \right) }
\end{displaymath}
for $L_{151} \ge L_{\rm break}$ and a flat power-law ($\alpha_{1}=0$)
for $L_{151} < L_{\rm break}$. This adds another free parameter to our
model as $\alpha_{1}$ has been replaced by $\alpha_{1} +
\alpha_{2}z$. However, 5 free parameters is still a small number
considering this is a 3-dimensional model: $\rho$ ($L,z$). The
redshift distribution is again specified by a two-tailed gaussian, as
in model B. The 2-D KS test was applied to the best-fits to this model
and gave $P_{\rm KS}=0.4-0.7$ for all the cosmological
models. Monte-Carlo simulations using the model G RLF give $L-z$
planes indistinguishable from the true data. In addition, this model
also fits the source counts at low fluxes well, giving a quasar
fraction at 0.1 Jy of $\approx 30\%$.  Similarly, we tested a variant
of model G with a one-tailed gaussian redshift distribution (i.e. no
decline in redshift for $z>z_{\circ}$).  This model (H) again fits
well with KS probabilities in the range $0.4-0.9$ and good Monte-Carlo
simulations and source counts fits. The likelihoods for model H are
slightly higher than for model G.

Comparing the relative likelihoods for models C and H, we find that
for cosmology I model H has a 10 times greater likelihood than C,
whereas for cosmologies II and III the likelihoods are comparable.
Since H is a 5 parameter model, whereas C has only 4 parameters, we
should only accept H as a best-fitting model if it can be shown that
it is a significantly better fit than C. To do this we need the ratio
of the posterior probabilities for each model confronted with the
data. The required ratio can be fairly approximated by the ratio of
the likelihoods multiplied by an `Ockham factor' ($< 1$) which
penalises the extra latitude afforded to model H by an extra free
parameter (e.g. Sivia 1996). For cosmologies II and III there is
clearly no motivation to prefer model H (since their likelihood ratios
are about 1). For cosmology I the ratio of the posterior probabilities
is at most 10, so although this hints that model H is to be preferred,
uncertainty in the value of the Ockham factor ensures this remains a
hint rather than a firm conclusion. The simulated $L_{151}-z$ data and
source counts plots for both models are similar and fit the data well.
Note that there are no 3CRR RLQs with $z>2.02$, despite the fact that
the highest redshift RLQ, 3C9, is luminous enough to be seen in the
3CRR sample at $z > 3$ and 3C191 at $z \approx 2.7$. In contrast,
there are nine 7C RLQs in the range $2<z<3$. This difference is
reproduced slightly better by model H than model C, due to the
steepening of the RLF at high-$z$. An additional check on the validity
of introducing an extra parameter comes from the size of its
error. For model H, the extra parameter has the value $\alpha_{2}=
0.46^{+0.19}_ {-0.29}$. These errors are approximately equivalent to
1$\sigma$ errors, so for the additional term to become zero (and hence
revert to a 4 parameter model) would only be a 1.6$\sigma$
result. Therefore, we conclude that we do not have sufficient evidence
for the 5 parameter models to be considered a better fit than the 4
parameter models and hence model C remains our preferred model until
further data suggest otherwise.

\subsection{Results} 

The RLF models C and H are plotted in Figure \ref{fig:qrlf} for
$z=0.5$, 1, 2 and 3 (for cosmology I), along with the binned RLF of
Section 4.1. Although at first glance, the binned and model RLFs
appear quite different, these apparent differences are easily
explained by the fact that the mean redshift varies from bin to bin,
as discussed in Section 4.1. The fairly flat slope of the low-$z$
binned RLF is due to the mean redshift increasing from 0.5 to 1.0 as
luminosity increases. This can clearly be seen by comparing the binned
points with the model lines for $z=0.5$ and $z=1.0$. At high redshift
($1.2<z<3.0$), both model C and H lines slightly underestimate the
number density in several of the bins. This may simply be a
consequence of small number statistics, because the largest deviation
is only $\approx 2\sigma$. The small sample size used here means that
the RLF is only well-constrained in regions of the $L_{151}-z$ plane
populated by 3CRR and 7C sources and extrapolations to other ranges of
luminosity and redshift should be regarded with caution.

It is however, useful to consider extrapolation of the RLF to lower
radio luminosities than those found in this sample. It has long been
known that the radio luminosity distribution of quasars is bi-modal,
with the separation into radio-quiet and radio-loud quasars occurring
at $\log_{10}$($L_{151}$) $ \sim 25.5$ (c.f. Kellermann et al. 1989).
The 6C and 7C source counts show that a break in the power-law is
required at $\log_{10}$($L_{151}) \approx 27$, although we cannot
constrain the slope of the RLQ luminosity function on the low
luminosity side of this break or the exact luminosity of the
break. The paucity of quasars (only 2) in the combined 3CRR/7C sample
with $\log_{10} $($L_{151}$) $ \leq 26.5$ is consistent with a break
in the RLF near this luminosity. It should be noted however that 7 of
the BLRGs in 3CRR \& 7C have $\log_{10} $($L_{151}$) $ < 26.5$. In
fact, due to the radio-optical correlation to be discussed in Section
5, BLRGs often have low radio luminosities and may contribute
significantly to the quasar RLF at these radio luminosities. However,
even including low optical luminosity quasars (BLRGs), there must
still be a break in the RLQ RLF to be consistent with the 6C/7C source
counts.

The 6C complete sample of Eales (1985a,b) was selected at 151 MHz with
flux limits $2.00 \leq S_{151} \leq 3.93$ Jy and covers 0.102 sr
(using the revised selection criteria from Blundell et al. 1998a).
Hence it lies in between the 3CRR and 7C samples in the $L_{151}-z$
plane and can be used to check our RLF models. It contains only seven
quasars (Eales 1985b; unpublished data), none of which are promoted
into the sample by their core fluxes (Blundell et al. 1998a;
unpublished data) . Integrating our model C RLF (for cosmology I) over
the 6C flux limits and sky area, we find that there should be, on
average, 16 quasars in the sample. Assuming Poisson statistics,
$p(\leq7)=0.01$ given mean $\mu=16$. The result is similar for model H
with 15 quasars predicted in the 6C sample. It is therefore just
plausible that the low quasar fraction in this sample (as commented on
by Eales) is simply a statistical fluke.

The radio-loud quasar RLF derived here has some remarkably similar
features to the OLF and XLF. The increase in $\log_{10}\rho(z)$ from
$z \approx 0.5$ to $z\approx 2$ for model C is $\approx1.5$, compared
to $\approx 1.8$ for medium luminosity quasars in the XLF (Boyle et
al. 1994) and $\approx 1.5$ for medium luminosity quasars in the OLF
(Goldschmidt \& Miller 1998). Boyle et al. (1994) fit the quasar XLF
with a redshift distribution which peaks at $z=1.6$ and then remains
constant at higher redshifts, in accordance with our best-fitting RLF
models. Of course, a decline in the OLF beyond $z\approx 3$ is now
established (e.g. Warren, Hewett \& Osmer 1994; Schmidt, Schneider \&
Gunn 1995), although these optically-selected samples could be biased
if there is significant dust at high redshift. Flat-spectrum radio
quasars also show decreasing number densities beyond $z\approx 3$
(Shaver et al. 1998), a result which is insensitive to the effects of
dust. The high luminosity part of the OLF is known to steepen at high
redshift (Goldschmidt \& Miller 1998) and we find weak evidence that
this may be occurring also in the RLF (e.g. model H). These
similarities are important to pursue further because radio sources are
known to be short-lived compared to the Hubble time (typically
$10^{7}$ yr; Scheuer 1995), so the evolution must really be telling us
about the evolution of the birth-rate (see also Blundell et
al. 1998a), and by analogy the same for the OLF.

\section{The radio-optical correlation}

\subsection{Absolute $B$ magnitude calculation}

$K$-corrections in the $B$-band for high redshift objects can be very
large and uncertain. The multi-colour photometry presented here avoids
this potential source of error, because it allows the luminosity of an
object in the rest-frame $B$-band to be calculated. For example, at $z
\sim 3$ rest-frame $B$-band emission is observed at $1.8 \mu$m (close
to the near-IR $K$-band). For sources at $z>0.5$, emitted $B$-band
radiation is observed between the $R$ and $K$ bands. Therefore, for
all 7C quasars a power-law continuum was fitted between the $R$ and
$K$ magnitudes and a rest-frame $B$-band luminosity calculated. This
was then converted to an absolute magnitude, $M_{B}$. The errors on
these values are typically $\pm 0.2$ mag; the main source of error
being the $B$ and $R$ magnitudes from the APM catalogue. Paper VIII
shows that this method provides a good estimate of $M_{B}$, even in
quasars reddened by $A_{\mathrm V} \approx 1$.

For the 3CRR quasars, complete $K$-band photometry has not been
published, so this approach was not possible. However, the maximum
redshift in 3CRR is $\sim 2$, so the $K$-correction terms are not so
large. For the 3CRR quasars a power-law optical continuum was
calculated using published $V$ and $R$ magnitudes or $V$ and $B$
magnitudes, where available. For the few sources with only a published
$V$ magnitude, an optical continuum slope of $\alpha_{\mathrm
opt}=0.5$ was assumed. This slope is equivalent to $B-K \approx 3$ and
is typical of unreddened quasars (Francis 1996). The calculated
absolute magnitudes are listed in Table B1 in Appendix B.
$K$-magnitudes were used (where available) in the calculation of the
absolute magnitudes of 3CRR radio galaxies referred to in Section 2.

\begin{figure}
\epsfxsize=0.5\textwidth
\hspace{-0.6cm}
\epsfbox{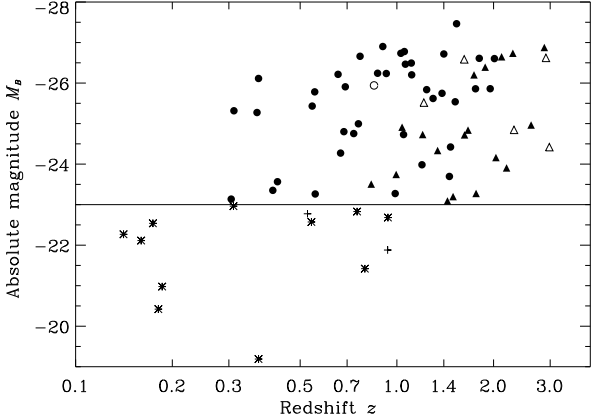}
{\caption[junk]{\label{fig:qsozmb} Absolute $B$ magnitude against
redshift for the 3CRR and 7C quasar samples. The plotting symbols are
the same as those used in Figure \ref{fig:qsopz}. BLRGs with $z\ge0.1$
are also plotted here (asterisks - 3CRR, plus signs - 7C). This plot
is for cosmology I.}}
\end{figure}

\begin{figure}
\epsfxsize=0.5\textwidth
\hspace{-0.6cm}
\epsfbox{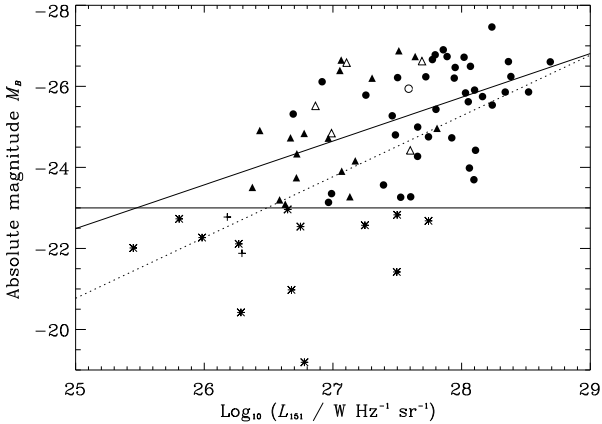}
{\caption[junk]{\label{fig:qsolrmb} Absolute B magnitude against
low-frequency radio luminosity for the 3CRR and 7C quasar samples.
The plotting symbols are the same as those used in Figure
\ref{fig:qsozmb}. Plotted also is the best-fit straight line to the
SSQs only, as described in the text (solid line). The dashed line is
the radio-optical correlation for SSQs of Serjeant et al. (1998)
transformed to cosmology I and $H_{\circ}=50~{\rm
km~s^{-1}Mpc^{-1}}$. This plot is for cosmology I.}}
\end{figure}

\subsection{Results}

All the numerical results in this Section are presented for cosmology
I. None of the conclusions differ significantly for either cosmology
II or III.  In Figure \ref{fig:qsozmb}, we plot the calculated
absolute magnitudes of the quasars and broad-line radio galaxies
against redshift. The horizontal line is the dividing line in absolute
magnitude between quasars and BLRGs from Section 3. The 5 CJSs are
considered separately from the SSQs in this section because of
possible orientation biases in their optical luminosities (e.g. Browne
\& Wright 1985). The BLRGs are also excluded because they are probably
not a complete sample. There is a very weak correlation between
$M_{B}$ and $z$ for the SSQs (in the sense that the quasars increase
in luminosity at greater $z$). Using Spearman's method of rank
correlation, the significance of the correlation between $M_{B}$ and
$z$ is 88\%. A radio-optical correlation is implicit in Figure
\ref{fig:qsozmb} in the redshift range ($1<z<2$) where there is the
most overlap between the 3CRR and 7C samples. It can clearly be seen
that the 3CRR quasars (which have radio luminosities $\sim 25$ times
greater than 7C quasars at the same $z$) are systematically brighter
than those in 7C. In this redshift range, the mean absolute magnitude,
$ \langle M_{B} \rangle $, for 3CRR SSQs is $-25.8$ with standard
error 0.2 mag. The 7C SSQs have $\langle M_{B} \rangle =-24.6 \pm 0.3$
mag, so are significantly fainter by about one magnitude. Note that
the BLRGs appear to be more tightly correlated than the quasars on
this plot. This is most likely due to the fact that all but 2 BLRGs
are from the 3CRR sample, so there is a tight $L_{151}-z$ correlation.

Figure \ref{fig:qsolrmb} shows the correlation between radio and
optical luminosities for the combined 3CRR and 7C SSQ samples. It can
clearly be seen that there is a much stronger correlation than in
Figure \ref{fig:qsozmb}. The significance of this correlation is
$>99.9\%$. This correlation does not appear to be caused by a
correlation between $M_{B}$ and $z$ (and a $L_{151}-z$ correlation),
because the statistical significance is very much greater. Also
plotted on Figure \ref{fig:qsolrmb} is the best-fit straight line to
the 3CRR/7C SSQs (calculated by minimising the sum of the squares of
the residuals). This line has a slope of $-2.5 \times (0.4 \pm 0.1)$,
i.e. $L_{\mathrm opt} \propto L_{\mathrm rad}^{0.4}$. The dispersion
about our straight-line fit is 1.0 magnitude. This is in fairly good
agreement with Serjeant et al. (1998) who found a slightly steeper
slope of $-2.5 \times (0.6 \pm 0.1)$, which is also plotted in
Fig. \ref{fig:qsolrmb} for comparison. One possible reason for our
finding a flatter slope than Serjeant et al. follows from their lack
of a limiting absolute magnitude in their definition of a
quasar. Their sample contains quasars with $M_{B} >-23$ (which we
would classify as BLRGs). These objects mostly have low radio
luminosities, therefore increasing the slope of the correlation. This
is evident in Fig. \ref{fig:qsolrmb}. There may be other systematic
differences between the RLQs in the Serjeant et al. MAQS sample and
those in the 7C sample. For example, the MAQS sample is selected at
408 MHz leading to a higher proportion of compact steep-spectrum radio
sources. We will consider these points in more detail in a future
paper.

To investigate whether the radio-optical correlation could be due to a
$M_{B}-z$ correlation, the Spearman partial rank correlation
coefficient was used. This assesses the statistical significance of
correlations between two variables in the presence of a third (Macklin
1982). For the correlation between $M_{B}$ and $z$ (independent of
$L_{151}$), $r_{M_{B}z,L_{151}} =-0.13$ with significance
$D_{M_{B}z,L_{151}}=-1.0$. The significance is equivalent to the
deviation from a unit variance normal distribution if there is no
correlation present. Hence the correlation between $M_{B}$ and $z$ is
marginally significant. Now we consider the correlation between
$M_{B}$ and $L_{151}$ (independent of $z$). We find
$r_{M_{B}L_{151},z} =-0.43$ and $D_{M_{B}L_{151},z} =-3.6$. Hence we
can be sure that the obvious radio-optical correlation of Figure
\ref{fig:qsolrmb} is not caused by the $M_{B}-z$ correlation which
plagues single flux-limited samples. The high radio luminosity but low
optical luminosity (but brighter than $M_{B}=-23$) quasar region of
Figure \ref{fig:qsolrmb} is genuinely devoid of quasars. There are no
selection effects which would cause a lack of quasars in this region.

Whenever a correlation such as this is presented, it is essential that
all possible sources of error, both random and systematic, are
considered. This is particularly important when combining samples with
different selection criteria. The 3CRR and 7C SSQ samples are however,
selected with very similar criteria (apart from their different flux
limits).  The only difference is their slightly different selection
frequencies; 3CRR is selected at 178 MHz and 7C at 151 MHz. Both are
optically complete, because all the sources in the 3CRR sample and the
7C Redshift Survey have been identified. Hence we have eliminated one
of the previously important possible sources of error, that of
`missing' quasars due to an optical magnitude limit.

There may be systematic errors between the two samples because the
optical data available for them is inhomogeneous. The right side of
Figure \ref{fig:qsolrmb} is composed mostly of 3CRR objects and the
left side mostly 7C, so could this explain the observed correlation?
The absolute magnitudes of the 7C quasars were calculated using
multi-colour photometry, so are likely to be good estimates of the
true values. Errors in $M_{B}$ due to reddening are not expected to be
crucial here, because $R$ and $K$ magnitudes were predominantly
used. Intrinsic reddening can be important when using $B$ magnitudes
for high redshift objects, because the observed radiation was emitted
in the UV. For the 3CRR quasars, $M_{B}$ was calculated using optical
photometry. Hence it is possible here that some quasars may be
significantly reddened and this is a possible source of error. In
addition, the magnitudes listed in LRL have not been corrected for
galactic extinction. However, both these effects would tend to cause
our calculated 3CRR $M_{B}$ values to be fainter than their true
values. This cannot explain the observed radio-optical correlation,
because the 3CRR quasars are systematically brighter than those in
7C. As noted above, the mean absolute magnitude of 3CRR SSQs is about
1 magnitude greater than the mean for 7C SSQs in the same redshift
range. Hence the 3CRR photometry must be systematically in error by
this amount for the correlation to disappear. It seems unlikely that
this could be the case. The numerical results presented in this
section are virtually independent of the choice of cosmology.

The data we have here is not detailed enough for us to explore the
nature of the radio-optical correlation in quasars. For a discussion
of possible physical causes of this effect, we refer the reader to
Serjeant et al. (1998) and Paper III. Note that the CJSs tend to
occupy positions on the optically bright side of the correlation. This
could be explained by orientation-dependent optical emission, however
the small number (6) of CJSs in the combined sample prevents us from
drawing any firm conclusions.

\section{Completeness of other samples of SSQs}

\begin{figure*}
\epsfxsize=0.9\textwidth
\begin{centering}
\epsfbox{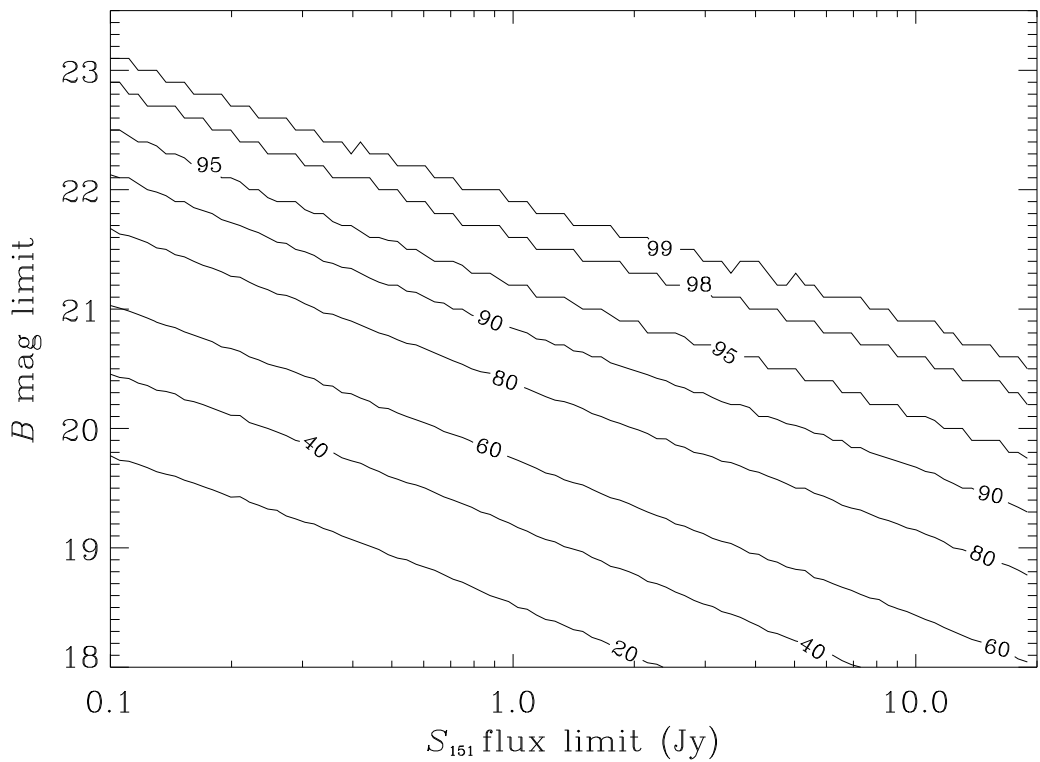}
\end{centering}
{\caption[junk]{\label{fig:mbslim} Percentage of quasars detected in
surveys as a function of their optical and radio flux limits.This plot
is for cosmology I.}}
\end{figure*}

Using the quasar radio luminosity function derived in the previous
section (model C), artificial RLQ samples were generated with a
Monte-Carlo simulation for a range of $S_{151}$ flux limits from 0.1
to 20 Jy. Each sample contains approximately $5 \times 10^{4}$
sources, each defined by $L_{151}$ and $z$. The straight-line fit to
the radio-optical correlation for SSQs derived in Section 5 defines a
characteristic value of $M_{B}$ for each value of $L_{151}$. The
scatter about this straight-line was assumed to be independent of
radio luminosity and a gaussian scatter of 1.0 magnitudes introduced
into the $M_{B}$ calculation. For each source, $M_{B}$ and $z$ were
then used to calculate its apparent $B$ magnitude (assuming
$\alpha_{\mathrm opt} =0.5$). Quasars with $M_{B}>-23$ (BLRGs) were
excluded from the simulated samples. The fraction of quasars brighter
than the limiting $B$ magnitude $B_{\mathrm lim}$ was then calculated
for each sample as a function of $B_{\mathrm lim}$.

Figure \ref{fig:mbslim} shows contours of the percentage of SSQs
expected to be detected in surveys as a function of their optical and
radio flux limits (for $\Omega_ {M}=1$, $\Omega_ {\Lambda}=0$). The
immediate conclusion to be drawn here is that as surveys go to fainter
radio fluxes, they must also go deeper optically, otherwise a
significant fraction of quasars are missed. For $\Omega_ {M}=0$,
$\Omega_ {\Lambda}=0$, the contours have steeper slopes across the
diagram, such that they diverge from the $\Omega_ {M}=1$, $\Omega_
{\Lambda}=0$ contours at faint radio fluxes to be 0.5 magnitudes in $B$ 
fainter at $S_{151}=0.1$ Jy. A similar behaviour is seen for $\Omega_
{M}=0.1$, $\Omega_ {\Lambda}=0.9$. It must be remembered that these
contours have been calculated assuming models for the radio-optical
correlation and the quasar RLF from the combined 3CRR/7C sample.
Therefore in regions not probed by these samples, it is an
extrapolation of the models. The area with the largest uncertainty is
therefore $S_{151}<0.5$ Jy. At $S_{151}=0.1$ Jy, typical errors on the
contours may be as large as $\pm 1.0$ in limiting $B$ magnitude.

According to Figure \ref{fig:mbslim}, 99\% of the 3CRR SSQ sample has
$B<20.5$ (i.e. brighter than the POSS limit). This is in good
agreement with the fact that all of them are detected on the POSS
plates. Our calculation predicts that 13 out of the 18 7C
steep-spectrum quasars have $B<20.5$, compared with the actual number
which is 11. Note that the completeness percentages calculated here
are strictly upper limits, because they do not take fully into account
the faintness of reddened quasars. The completeness contours derived
here only apply to low-frequency selected samples, e.g. 151 MHz. This
is because high-frequency samples contain a large fraction of FSQs
which are typically brighter than the mean SSQ radio-optical
correlation (Wills \& Lynds, 1978; Browne \& Wright, 1985). Therefore
FSQs are more likely to be brighter than the optical magnitude limits
of samples.

We can use Figure \ref{fig:mbslim} to estimate the completeness of
various low-frequency radio selected quasar samples. This allows us to
investigate possible bias in the sample properties due to selection
effects. The large 7C quasar sample of Riley et al. (1998) has a faint
radio flux limit of $S_{151}>0.1$ Jy and a bright optical limit of
$R<20$. This sample is probably only approximately 40\% complete.
There are other new radio quasar samples which use POSS-I ($R \approx
20$) identifications. The B3-VLA quasar sample (Vigotti et al. 1997)
with a flux limit $S_{408} \geq 0.1$ Jy is estimated to be 55\%
complete from our plot (c.f. their estimate based on optical magnitude
histograms of 70\%). For samples with $S_{151} \ltsimeq 1$ Jy, one
must use optical data significantly deeper than the POSS-I limit in
order to avoid severe incompleteness in these samples.

The MAQS sample of Serjeant et al. (1998) is selected by
$S_{408}>0.95$ Jy and $B<22.5$. We estimate that this sample is $99\%$
complete, due to its faint optical limit (and its relatively bright
radio limit). However, the higher selection frequency of 408 MHz
samples rest-frame $>1$ GHz emission for $z> 1.5$ quasars. At these
rest-frame frequencies cores and jets often dominate over extended
emission, so the extended radio luminosity is lower and hence the
quasars will be lower down the radio--optical correlation. It is
possible however that a beamed optical--UV component, as discussed in
the previous paragraph, may compensate in these objects. Therefore we
conclude that the MAQS sample is likely to have excluded very few
quasars due to its optical magnitude limit. A comparison of our RLF
models with the MAQS sample will appear elsewhere (Serjeant et al. in
preparation).

\section{Conclusions and summary}

The sample of 7C quasars presented here is {\it complete} because
every source in the 7C Redshift Survey has been identified. Therefore
we do not suffer from colour biases excluding red and/or faint
quasars.  However, if orientation-based unified schemes (e.g.
Antonucci 1993) are correct, then powerful radio galaxies are reddened
quasars. For intermediate values of reddening (A$_{V} \approx 2$),
different observers may classify objects as either radio galaxies or
quasars. These issues will be investigated further in Paper VIII. The
main conclusions to be drawn from this paper are:

\begin{itemize}
\item The radio luminosity function of radio-loud quasars has been
derived. The best-fitting model is a single power-law with slope
$\alpha_{1} = 1.9 \pm 0.1$. We find that there must be a break in the
RLQ RLF at ${\rm \log}_{10}(L_{151} /$ W Hz$^{-1}$ sr$^{-1}) \ltsimeq
27$, in order for the models to be consistent with the 7C and 6C
source counts. The $z$-dependence of the RLF follows a one-tailed
gaussian which peaks at $z=1.7 \pm 0.2$ and remains constant at higher
redshifts. Models with or without a redshift cut-off beyond $z\approx
2$ fit almost equally well. We find marginal evidence that the RLF
slope steepens as redshift increases. This form for the evolving RLF
is very similar to the OLF determined from optically-selected samples
(e.g. Goldschmidt \& Miller 1998).

\item We have confirmed the existence of a significant correlation
between the low-frequency radio and optical luminosities of SSQs, as
reported in Serjeant et al. (1998). This correlation is not due to
correlations with redshift or sample selection effects. 
 
\item Using the derived radio-loud quasar RLF and the observed
radio-optical correlation, we have been able to estimate the fraction
of quasars identified in low-frequency radio-selected samples with
various magnitude limits. It is found that samples with faint radio
flux limits must also go much deeper in the optical than the POSS-I
plates ($R \approx 20$), or there will be serious incompleteness in
the samples. This would introduce many biases into the analysis of
such samples, because only the optically brightest quasars would be
included.

\end{itemize}

\section*{Acknowledgements}
Special thanks to Steve Eales, Gary Hill, Julia Riley and David
Rossitter for help with various aspects of the 7C Redshift Survey. We
warmly thank the referee, Pippa Goldschmidt, for her constructive
comments. We would like to thank Andrew Bunker and Isobel Hook for
obtaining a spectrum of 5C7.95 for us and Richard Saunders for help
with the spectroscopy of 5C6.95. We thank the staff at the WHT and
UKIRT for technical support. The United Kingdom Infrared Telescope is
operated by the Joint Astronomy Centre on behalf of the U.K. Particle
Physics and Astronomy Research Council. We acknowledge the UKIRT
Service Programme for some of the $K$-band imaging. The William
Herschel Telescope is operated on the island of La Palma by the Isaac
Newton Group in the Spanish Observatorio del Roque de los Muchachos of
the Instituto de Astrofisica de Canarias.  This research has made use
of the NASA/IPAC Extra-galactic Database, which is operated by the Jet
Propulsion Laboratory, Caltech, under contract with the National
Aeronautics and Space Administration. CJW thanks PPARC for receipt of
a studentship.

\clearpage

\appendix

\section{Optical spectra}

\begin{figure*}
\epsfxsize=0.95\textwidth
\hspace{1.25cm}
\begin{centering}
\epsfbox{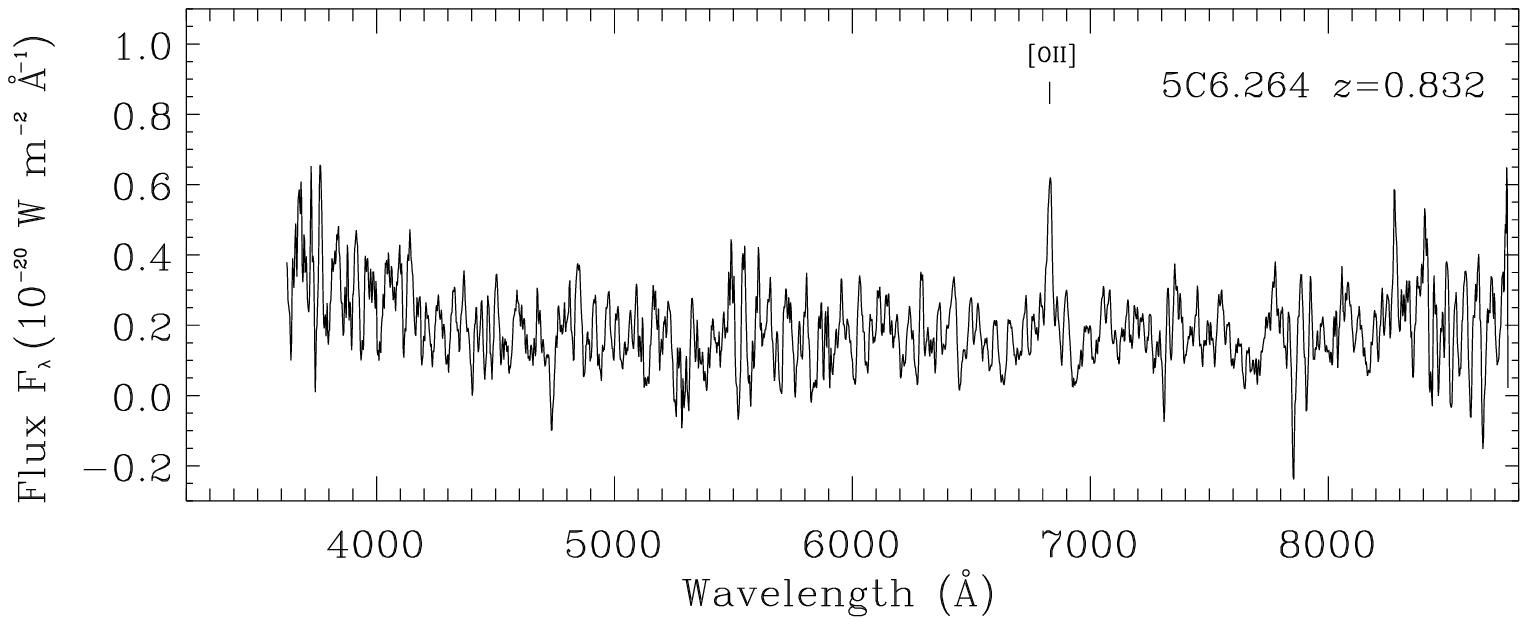} 
\end{centering}
\vspace{1.0cm} {\caption[junk]{\label{fig:spec6p264} Optical spectrum
of the quasar 5C6.264. This object was observed through cloud and the
spectrum suffered several magnitudes of extinction. The spectrum has a
blue continuum and a single narrow emission line which we identify as
[OII] $\lambda$3727 at a redshift of 0.832. The $K$-band image of this
object shows a very strong unresolved component and is $\sim 1$ mag
brighter than the mean $K-z$ relation for radio galaxies (Eales et
al. 1997). An infrared spectrum of 5C6.264 shows a broad H$\alpha$
line (Paper VIII). We attribute the lack of broad lines in the optical
spectrum presented here (MgII $\lambda 2799$ expected at 5130\AA)
as solely due to its poor snr. }}
\end{figure*}

\begin{figure*}
\epsfxsize=0.95\textwidth
\hspace{1.25cm}
\begin{centering}
\epsfbox{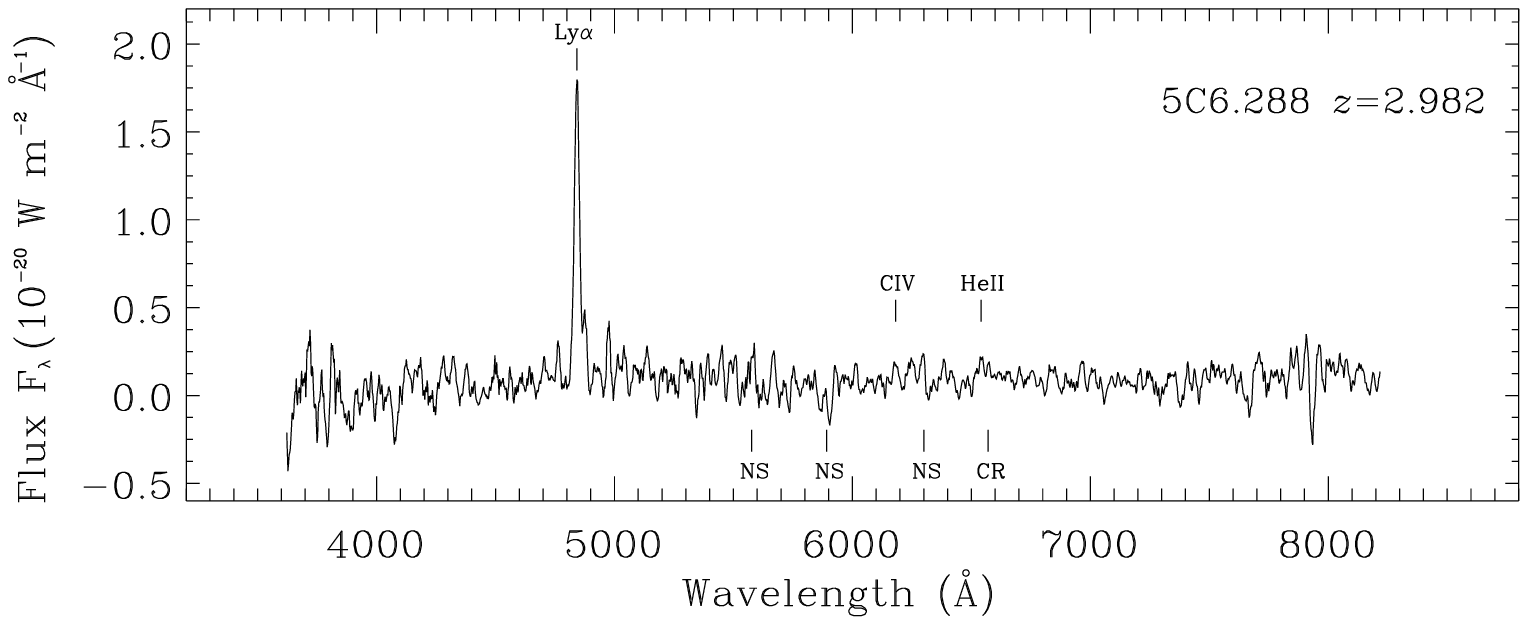} 
\end{centering}
\vspace{1.0cm} {\caption[junk]{\label{fig:spec6p288} Optical spectrum
of the probable quasar 5C6.288. This spectrum shows strong narrow
Lyman-$\alpha$ emission and weak narrow CIV and HeII lines. The CIV
and HeII lines are clearly visible on the 2-dimensional spectrum. The
CIII] $\lambda 1909$ line is not observed as it falls in the
atmospheric absorption band at $\sim 7600$\AA. Regions of the spectrum
affected by night sky lines and cosmic rays are marked NS and CR,
respectively. The bright absolute magnitude ($M_{B}=-24.4$) and
compact nature at $K$-band are suggestive of a lightly reddened quasar
($A_{\mathrm V}\sim 1$), which explains the lack of broad UV lines in
the optical spectrum (see Paper VIII).}}
\end{figure*}

\begin{figure*}
\epsfxsize=0.9\textwidth
\hspace{0.25cm}
\begin{centering}
\epsfbox{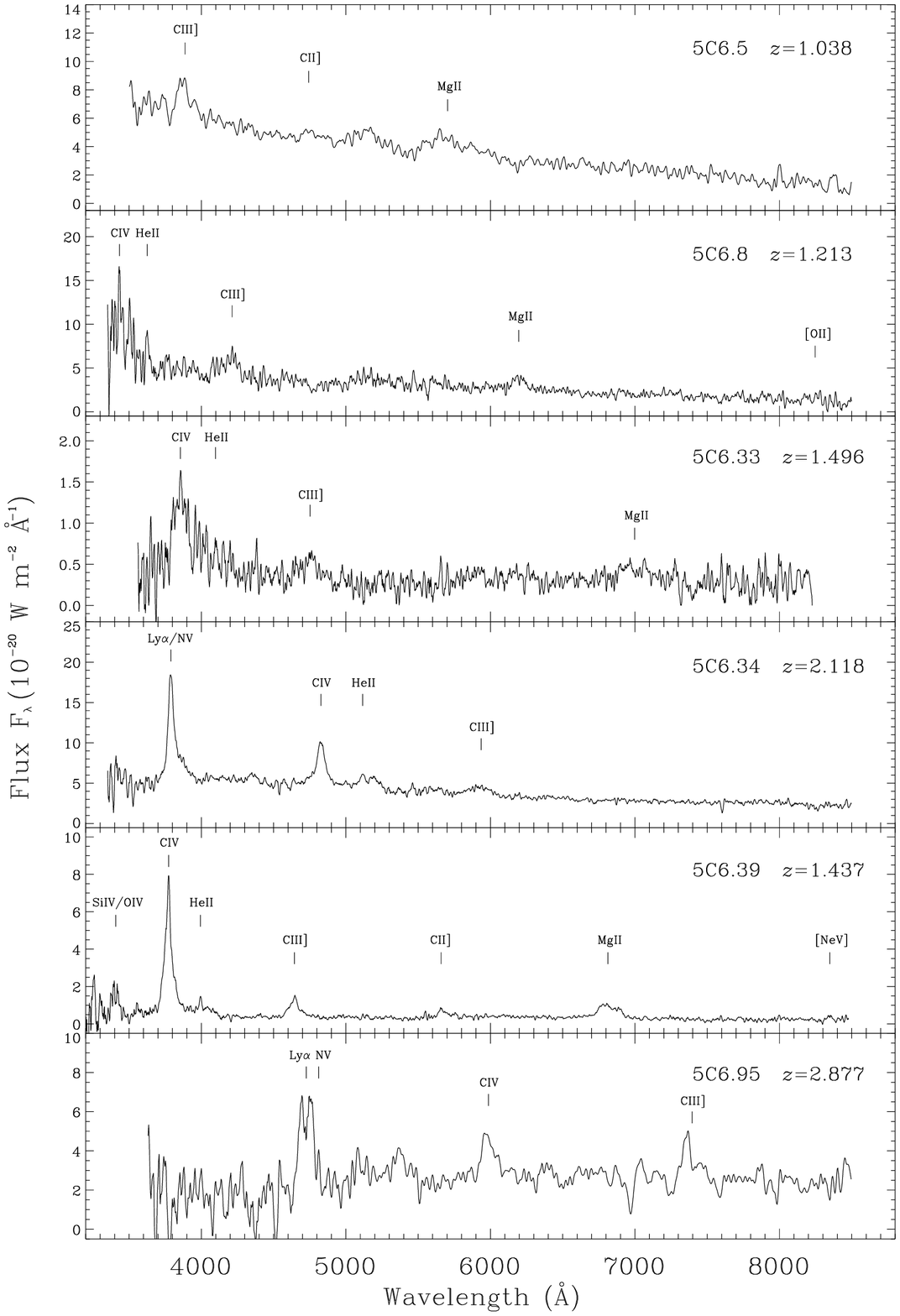} 
\end{centering}
\vspace{1.0cm}
{\caption[junk]{\label{fig:spec} 
}}
\end{figure*}

In this appendix we present the optical spectra and emission line data
for the 7C Redshift Survey quasars and broad-lined radio galaxies.
Figures \ref{fig:spec6p264} and \ref{fig:spec6p288} are the optical
spectra of 5C6.264 and 5C6.288, respectively. These are the two
quasars with no evidence of broad emission lines which will be
discussed further in Paper VIII. Figure \ref{fig:spec} displays the
optical spectra for the other 24 7C quasars and broad-line radio
galaxies. Prominent emission lines are labelled. Table A1 lists the
properties of the emission lines measured from all these spectra.


\begin{table*}
\footnotesize
\begin{center}
\begin{tabular}{lrlccrrrrl}
\hline\hline
\mc{1}{c}{name} &\mc{1}{c}{$z$} &\mc{1}{l}{line} &\mc{1}{c}{$\lambda_{\mathrm rest}$} &\mc{1}{c}{$\lambda_{\mathrm obs}$} &\mc{1}{c}{FWHM}           &\mc{1}{c}{flux} &\mc{1}{c}{snr}   &\mc{1}{c}{W$_{\lambda}$}   &\mc{1}{l}{notes} \\  
\mc{1}{c}{} &\mc{1}{c}{ } &\mc{1}{c}{ }    &\mc{1}{c}{(\AA)}  &\mc{1}{c}{(\AA)} &\mc{1}{c}{(\AA)}  &\mc{1}{c}{(W m$^{-2}$)} &\mc{1}{c}{$\sigma$} &\mc{1}{c}{(\AA)} &\mc{1}{c}{ }\\
\hline\hline
  5C6.5      &  1.038 &  C III]            & 1909  & 3887 $\pm$ 15   &  68 $\pm$ 10   & 1.2e-17  &  5    &  71 $\pm$ \f 20 &  abs                    \\ 
 7C B020414.4+332221 &  $\pm$ &  C II]     & 2326  & 4743 $\pm$ 10   &  58 $\pm$ 12   & 2.6e-18  &  2.1  &  19 $\pm$ \f \f 5 &                         \\
             &   .002 &  Mg II             & 2799  & 5704 $\pm$ \f 2 & 139 $\pm$ 40   & 1.3e-17  &  4    & 138 $\pm$ \f 40 &                \\ \hline
  5C6.8      &  1.213 &  C IV              & 1549  & 3434 $\pm$ \f 7 &  34 $\pm$ 20   & 1.6e-17  &  2.8  &  65 $\pm$ \f 35 &  in noise		      \\
 7C B020439.6+313752 &  $\pm$ &  He II     & 1640  & 3626 $\pm$ \f 3 &  23 $\pm$ \f 6 & 3.4e-18  &  1.6  &  16 $\pm$ \f \f 8 &                 	      \\
             &   .002 &  C III]            & 1909  & 4213 $\pm$ \f 5 & 110 $\pm$ 30   & 1.4e-17  &  5    &  81 $\pm$ \f 30 & 			      \\
             &        &  Mg II             & 2799  & 6196 $\pm$ \f 5 & 101 $\pm$ 20   & 1.0e-17  &  6    &  71 $\pm$ \f 16 &  abs      		      \\
             &        &  [O II]            & 3727  & 8247 $\pm$ \f 5 &  24 $\pm$ \f 5 & 1.0e-18  &  1.3  &  23 $\pm$ \f \f 8 & 		       \\ \hline
  5C6.33     &  1.496 &  C IV              & 1549  & 3855 $\pm$ 10   & 150 $\pm$ 40   & 1.9e-18  &  7    & 220 $\pm$ \f 70 &  abs     		      \\
7C B020836.3+333419  &  $\pm$ &  He II     & 1640  & 4098 $\pm$ 10   & 110 $\pm$ 30   & 4.0e-19  &  2.3  &  53 $\pm$ \f 17 &  			      \\
             &   .003 &  C III]            & 1909  & 4752 $\pm$ 20   & 170 $\pm$ 40   & 9.5e-19  &  4    & 229 $\pm$ \f 70 &  abs          	      \\
             &        &  Mg II             & 2799  & 6998 $\pm$ 20   & 209 $\pm$ 50   & 8.8e-19  &  4    & 200 $\pm$ \f 60 &  abs                  \\ \hline
5C6.34       &  2.118 &  Ly $\alpha$ / N V & 1216  & 3790 $\pm$ \f 6 &  47 $\pm$ 10   & 5.3e-17  &  35   & 198 $\pm$ \f 40 &  abs     		      \\
7C B020838.5+312131  &  $\pm$ &  C IV      & 1549  & 4827 $\pm$ \f 3 &  76 $\pm$ 10   & 1.5e-17  &  15   &  71 $\pm$ \f 10 &  abs      		      \\
             &   .002 &  He II             & 1640  & 5116 $\pm$ \f 4 &  42 $\pm$ \f 8 & 6.1e-18  &  6    &  38 $\pm$ \f 12 &                 	      \\
             &        &  C III]            & 1909  & 5936 $\pm$ 15   & 146 $\pm$ 20   & 8.1e-18  &  6    &  56 $\pm$ \f 30 &  cr in line \\ \hline
 5C6.39      &  1.437 &  Si IV / O IV      & 1400  & 3408 $\pm$ 15   &  57 $\pm$ 20   & 8.9e-19  &  7    & 193 $\pm$ 120 & 		              \\
7C B020853.8+321446  &  $\pm$ &  C IV      & 1549  & 3774 $\pm$ \f 3 &  65 $\pm$ 12   & 6.1e-18  &  70   & 415 $\pm$ \f 80 & 		              \\
             &   .003 &  He II             & 1640  & 3994 $\pm$ \f 3 &  17 $\pm$ \f 7 & 3.5e-19  &  6    &  47 $\pm$ \f 40 &  poss. broad base	      \\
             &        &  C III]            & 1909  & 4644 $\pm$ \f 5 &  89 $\pm$ 19   & 1.2e-18  &  9    & 220 $\pm$ \f 35 &  abs	              \\
             &        &  C II]             & 2326  & 5659 $\pm$ 12   &  45 $\pm$ 10   & 3.6e-19  &  5    &  71 $\pm$ \f 28 & 		              \\
             &        &  Mg II             & 2799  & 6813 $\pm$ 17   & 146 $\pm$ 25   & 1.9e-18  &  15   & 475 $\pm$ 110 &  abs and atm abs 	      \\
             &        &  [Ne V]            & 3426  & 8348 $\pm$ \f 5 &  50 $\pm$ 25   & 1.8e-19  &  4    &  81 $\pm$ \f 46 & 		        \\ \hline
  5C6.95     &  2.877 &  Ly $\alpha$       & 1216  & 4726 $\pm$ 13   & 109 $\pm$ \f 9 & 5.6e-18  &  14   & 375 $\pm$ \f 90 &  abs                    \\
7C B021149.2+293640  &  $\pm$ &  N V       & 1240  & 4812 $\pm$ \f 3 &  21 $\pm$ \f 4 & 7.1e-19  &  1.8  &  60 $\pm$ \f 12 &                         \\
             &   .007 &  C IV              & 1549  & 5986 $\pm$ 20   &  92 $\pm$ 11   & 2.4e-18  &  14   &  95 $\pm$ \f 35 &  abs          	      \\
             &        &  C III]            & 1909  & 7396 $\pm$ 25   & 128 $\pm$ 24   & 2.7e-18  &  14   & 117 $\pm$ \f 40 &  abs          	 \\ \hline
5C6.160      &  1.624 &  C IV              & 1549  & 4065 $\pm$ \f 5 &  66 $\pm$ \f 8 & 3.2e-17  &  13   & 291 $\pm$ \f 45 &  abs                       \\
 7C B021444.8+332114 &  $\pm$ &  He II     & 1640  & 4304 $\pm$ \f 5 &  53 $\pm$ 12   & 2.7e-18  &  1.7  &  31 $\pm$ \f 13 &  abs                    \\
             &   .002 &  C III]            & 1909  & 5004 $\pm$ \f 7 &  78 $\pm$ 20   & 1.1e-17  &  6    & 165 $\pm$ \f 45 &  abs      		      \\
             &        &  Mg II             & 2799  & 7351 $\pm$ 20   & 140 $\pm$ 60   & 8.0e-18  &  1.5  & 201 $\pm$ \f 50 &  abs      	      \\ \hline
5C6.237      &  1.620 &  C IV              & 1549  & 4058 $\pm$ 15   &  70 $\pm$ 20   & 2.6e-17  &  12   &  78 $\pm$ \f 28 &  abs \& cr in line	      \\
 7C B021749.9+322724 &  $\pm$ &  He II     & 1640  & 4280 $\pm$ 20   &  86 $\pm$ 25   & 1.0e-17  &  4    &  35 $\pm$ \f \f 7 &                 	      \\
             &   .003 &  C III]            & 1909  & 5007 $\pm$ 15   & 102 $\pm$ 28   & 1.4e-17  &  5    &  52 $\pm$ \f 12 &  abs     		      \\
             &        &  Mg II             & 2799  & 7332 $\pm$ \f 8 &  82 $\pm$ 24   & 4.0e-18  &  2.6  &  27 $\pm$ \f 10 &  abs \& cr in line         \\ \hline
5C6.251      &  1.665 &  Si IV / O IV      & 1400  & 3726 $\pm$ \f 8 &  44 $\pm$ \f 8 & 5.8e-18  &  5    &  47 $\pm$ \f 18 &                 	      \\
7C B021850.4+340239  &  $\pm$ &  C IV      & 1549  & 4107 $\pm$ 12   &  91 $\pm$ 10   & 1.2e-17  &  8    & 108 $\pm$ \f 15 &  abs      		      \\
             &   .002 &  He II             & 1640  & 4376 $\pm$ 25   & 110 $\pm$ 60   & 3.2e-18  &  1.2  &  30 $\pm$ \f 15 & 		              \\
             &        &  C III]            & 1909  & 5060 $\pm$ 20   & 114 $\pm$ 40   & 7.2e-18  &  3.4  &  85 $\pm$ \f 24 &         		      \\
             &        &  C II]             & 2326  & 6199 $\pm$ \f 6 &  72 $\pm$ 28   & 1.0e-18  &  1.3  &  21 $\pm$ \f \f 9 & 		              \\
             &        &  [Ne IV]           & 2424  & 6463 $\pm$ \f 4 &  20 $\pm$ \f 4 & 2.1e-19  &  1.1  &   4 $\pm$ \f \f 3 & 		              \\
             &        &  Mg II             & 2799  & 7410 $\pm$ 40   & 155 $\pm$ 25   & 5.2e-18  &  4    & 131 $\pm$ \f 50 &  atm abs            \\ \hline
5C6.264      &  0.832 &  [OII]             & 3727  & 6829 $\pm$ \f 2 &  23 $\pm$ \f 2 & 1.8e-19  &  8    &  62 $\pm$ \f 15 &                 	      \\
7C B021949.1+334341  &  $\pm$ &            &       &                 &                &          &       &               &           		      \\
             &   .001 &                    &       &                 &                &          &       &               & 	              \\ \hline
5C6.282      &  2.195 &  Ly $\alpha$       & 1216  & 3902 $\pm$ 10   &  17 $\pm$ \f 5 & 3.2e-19  &  2.9  &  62 $\pm$ \f 31 &  abs		              \\
7C B022140.8+340609&  $\pm$ &  N V         & 1240  & 3965 $\pm$ \f 5 &  43 $\pm$ 12   & 4.2e-19  &  4    &  85 $\pm$ \f 32 &           	      \\
             &   .003 &  C IV              & 1549  & 4950 $\pm$ 20   & 150 $\pm$ 40   & 3.5e-19  &  3.5  &  71 $\pm$ \f 32 &  abs     		      \\
             &        &  He II             & 1640  & 5235 $\pm$ \f 4 &  80 $\pm$ 32   & 1.7e-19  &  2.9  &  27 $\pm$ \f 16 &  abs     		      \\
             &        &  C III]            & 1909  & 6063 $\pm$ 30   & 123 $\pm$ 40   & 1.4e-19  &  1.8  &  31 $\pm$ \f 11 & 		        \\ \hline
5C6.286      &  1.339 &  C IV              & 1549  & 3625 $\pm$ \f 7 &  74 $\pm$ \f 6 & 2.9e-17  &  7    & 400 $\pm$ 100 &  abs     	              \\
 7C B022209.4+314550 &  $\pm$ &  C III]    & 1909  & 4451 $\pm$ 10   &  86 $\pm$ 15   & 6.3e-18  &  2.6  & 121 $\pm$ \f 38 &                  	      \\
             &   .003 &  Mg II             & 2799  & 6563 $\pm$ 20   & 169 $\pm$ 25   & 8.8e-18  &  3.3  & 460 $\pm$ 100 &  abs     	   \\ 
\hline\hline         
\end{tabular}
{\caption[Table of observations]{\label{tab:emli1} 
}}
\normalsize
\end{center}              
\end{table*}

\addtocounter{table}{-1}

\begin{table*}
\footnotesize
\begin{center}
\begin{tabular}{lrlccrrrrl}
\hline\hline
\mc{1}{c}{name} &\mc{1}{c}{$z$} &\mc{1}{l}{line} &\mc{1}{c}{$\lambda_{\mathrm rest}$} &\mc{1}{c}{$\lambda_{\mathrm obs}$} &\mc{1}{c}{FWHM}           &\mc{1}{c}{flux}  &\mc{1}{c}{snr} &\mc{1}{c}{W$_{\lambda}$}   &\mc{1}{l}{notes} \\  
\mc{1}{c}{} &\mc{1}{c}{ } &\mc{1}{c}{ }    &\mc{1}{c}{(\AA)}  &\mc{1}{c}{(\AA)} &\mc{1}{c}{(\AA)}  &\mc{1}{c}{(W m$^{-2}$)} &\mc{1}{c}{$\sigma$} &\mc{1}{c}{(\AA)} &\mc{1}{c}{ }\\
\hline\hline
5C6.287      &  2.296 &  Ly $\alpha$ / N V & 1216  & 4005 $\pm$ \f 5   &  88 $\pm$ 12 & 8.5e-17  &  15   & 340 $\pm$ 100 &                 	      \\
7C B022220.0+313901  &  $\pm$ &  C IV      & 1549  & 5114 $\pm$   15   & 101 $\pm$ 18 & 3.5e-17  &  8    & 192 $\pm$ \f 80 &  abs		              \\
             &   .002 &  He II             & 1640  & 5408 $\pm$ \f 8   &  35 $\pm$ 30 & 8.5e-18  &  1.4  &  57 $\pm$ \f 35 & 		              \\
             &        &  C III]            & 1909  & 6288 $\pm$   12   &  80 $\pm$ 40 & 3.7e-18  &  1.9  & 145 $\pm$ \f 25 &               \\ \hline
5C6.288      &  2.982 &  Ly $\alpha$       & 1216  & 4842 $\pm$ \f 5   &  23 $\pm$ \f 3 & 6.1e-19  &  12   & 240 $\pm$ 100 &                 	      \\
7C B022219.3+310527  &  $\pm$ &  C IV      & 1549  & 6181 $\pm$ \f 6   &  22 $\pm$ \f 5 & 2.7e-20  &  1.4  &  36 $\pm$ \f 15 &     		              \\
             &   .003 &  He II             & 1640  & 6541 $\pm$ \f 6   &  24 $\pm$ \f 6 & 4.2e-20  &  1.6  &  34 $\pm$ \f 12 & 		        \\ \hline
5C6.291      &  2.910 &  Ly $\alpha$ / N V & 1216  & 4752 $\pm$ \f 1   &  22 $\pm$ \f 2 & 2.9e-18  &  90   & 372 $\pm$ \f 60 &  abs			      \\
7C B022309.7+340801  &  $\pm$ &C IV        & 1549  & 6022 $\pm$ \f 2   &  25 $\pm$ \f 5 & 1.7e-19  &  12   &  14 $\pm$ \f \f 8 &  abs                       \\
             &   .002 &  He II             & 1640  & 6417 $\pm$ \f 2   &  30 $\pm$ \f 4 & 2.2e-19  &  9    &  18 $\pm$ \f \f 7 &  		              \\
             &        &  C III]            & 1909  & 7465 $\pm$ \f 2   & 145 $\pm$ 25 & 7.6e-19  &  17   &  54 $\pm$ \f 11 &  narrow core    \\ \hline
7C0808+2854  &  1.883 &  Ly $\alpha$       & 1216  & 3518 $\pm$   12   & 105 $\pm$ 20 & 5.8e-17  &  50   & 190 $\pm$ \f 45 &  abs		        \\
7C B080832.1+285402 &  $\pm$ &Si IV / O IV & 1400  & 4020 $\pm$   10   & 117 $\pm$ 25 & 5.2e-18  &  6    &  22 $\pm$ \f \f 9 &  abs     \\
             &   .003 &  C IV              & 1549  & 4466 $\pm$ \f 9   & 108 $\pm$ 20 & 1.9e-17  &  15   &  90 $\pm$ \f 18 &  abs			      \\
             &        &  C III]            & 1909  & 5485 $\pm$   10   & 130 $\pm$ 25 & 7.0e-18  &  8    &  46 $\pm$ \f 11 &  		              \\
             &        &  Mg II             & 2799  & 8070 $\pm$ \f 4   & 200 $\pm$ 40 & 2.1e-18  &  3.3  &  24 $\pm$ \f 11 &  abs	 	        \\ \hline
5C7.17       &  0.936 &  C III]            & 1909  & 3695 $\pm$   15   &  95 $\pm$ 25 & 1.8e-18  &  9    & 302 $\pm$ 140 &  abs			      \\
7C B080956.4+270058  &  $\pm$ &  Mg II     & 2799  & 5407 $\pm$   15   & 202 $\pm$ 60 & 1.4e-18  &  8    & 260 $\pm$ 120 & 			      \\
             &   .001 &  [Ne V]            & 3346  & 6479 $\pm$ \f 2   &  14 $\pm$ \f 4 & 9.6e-20  &  1.8  &  27 $\pm$ \f \f 9 & 			      \\
             &        &  [O II]            & 3727  & 7217 $\pm$ \f 3   &  23 $\pm$ \f 3 & 3.5e-19  &  9    &  76 $\pm$ \f 18 & 			      \\
             &        &  [Ne III]          & 3869  & 7488 $\pm$ \f 2   &  16 $\pm$ \f 3 & 9.9e-20  &  2.5  &  24 $\pm$ \f \f 7 & 		  \\ \hline
5C7.70       &  2.617 &  Ly $\alpha$       & 1216  & 4420 $\pm$ \f 4   &  40 $\pm$ 10 & 3.9e-19  &  8    &  88 $\pm$ \f 37 &  abs			      \\
7C B081400.9+292731&  $\pm$ &  C IV        & 1549  & 5633 $\pm$   12   & 100 $\pm$ 50 & 1.9e-19  &  2.6  &  37 $\pm$ \f 25 &  abs			      \\
             &   .003 &  He II             & 1640  & 5947 $\pm$   14   & 140 $\pm$ 70 & 1.8e-19  &  2.2  &  37 $\pm$ \f 25 &    		      \\
             &        &  C III]            & 1909  & 6905 $\pm$ \f 5   & 100 $\pm$ 50 & 4.2e-19  &  8    &  95 $\pm$ \f 29 &  abs	        \\ \hline
5C7.85       &  0.995 &  C III]            & 1909  & 3799 $\pm$ \f 7   &  70 $\pm$ 30 & 9.3e-19  &  6    &  43 $\pm$ \f 20 &   cr in line	      \\
7C B081445.6+272601  &  $\pm$ &  Mg II     & 2799  & 5602 $\pm$   12   & 128 $\pm$ 20 & 2.1e-18  &  10   &  77 $\pm$ \f 22 &  abs	         	      \\
             &   .001 &  [O II]            & 3727  & 7436 $\pm$ \f 3   &  29 $\pm$ \f 3 & 2.6e-19  &  3.6  &  11 $\pm$ \f \f 3 &                   	      \\
             &        &  [Ne III]          & 3869  & 7718 $\pm$ \f 5   &  29 $\pm$ \f 3 & 2.9e-19  &  3.8  &  14 $\pm$ \f \f 3 &            	      \\ \hline
5C7.87       &  1.764 &  Ly $\alpha$       & 1216  & 3370 $\pm$ \f 4   &  28 $\pm$ 16 & 3.0e-18  &  2.7  &  96 $\pm$ \f 60 &   in noise	      \\
7C B081503.9+245827 &  $\pm$ &  C IV       & 1549  & 4278 $\pm$ \f 8   &  71 $\pm$ 15 & 4.2e-18  &  30   & 373 $\pm$ \f 74 &   abs     	      \\
             &   .002 &  He II             & 1640  & 4534 $\pm$ \f 5   &  69 $\pm$ 23 & 4.5e-19  &  6    &  41 $\pm$ \f 15 &  		              \\
             &        &  C III]            & 1909  & 5263 $\pm$   10   &  52 $\pm$ 30 & 9.1e-19  &  9    & 148 $\pm$ \f 38 &  	        	      \\
             &        &  Mg II             & 2799  & 7742 $\pm$   10   &  91 $\pm$ 30 & 5.8e-19  &  3.3  &  80 $\pm$ \f 40 &   in atm abs          \\ \hline
 5C7.95      &  1.203 &  C IV              & 1549  & 3413 $\pm$ \f 5   &  84 $\pm$ 24 & 1.7e-17  &  21   & 226 $\pm$ \f 57 &   abs on blue wing      \\
7C B081520.0+244506  &  $\pm$ &  C III]    & 1909  & 4195 $\pm$ \f 8   &  78 $\pm$ 22 & 4.3e-18  &  14   &  75 $\pm$ \f 12 &  		        \\
             &   .002 &  Mg II             & 2799  & 6171 $\pm$   10   & 150 $\pm$ 45 & 9.5e-18  &  16   & 194 $\pm$ \f 46 &  	              \\
             &        &  [O II]            & 3727  & 8213 $\pm$   12   &  40 $\pm$ 20 & 7.9e-19  &  3.1  &  22 $\pm$ \f 10 & 	             \\ \hline
5C7.118      &  0.527 &  Mg II             & 2799  & 4293 $\pm$ \f 3   &  59 $\pm$ 15 & 7.5e-18  &  31   &  65 $\pm$ \f 14 &   cr in line	      \\
7C B081615.1+265130 &  $\pm$ &  [Ne V]     & 3346  & 5109 $\pm$ \f 2   &  19 $\pm$ \f 4 & 8.8e-20  &  1.4  &   2 $\pm$ \f \f 1 & 		              \\
             &   .002 &  [Ne V]            & 3426  & 5230 $\pm$ \f 2   &  15 $\pm$ \f 3 & 9.0e-19  &  14   &  14 $\pm$ \f \f 5 &   Ca II 3933 abs ?          \\
             &        &  H $\gamma$        & 4340  & 6638 $\pm$   12   & 115 $\pm$ 20 & 1.2e-18  &  5    &  29 $\pm$  \f\ f 7 &       		      \\
             &        &  H $\beta$         & 4861  & 7424 $\pm$ \f 8   & 120 $\pm$ 18 & 3.4e-18  &  9    &  89 $\pm$ \f 17 & 		              \\
             &        &  [O III]           & 4959  & 7568 $\pm$ \f 3   &  27 $\pm$ 12 & 2.1e-19  &  2.5  &   5 $\pm$ \f \f 2 & 		              \\
             &        &  [O III]           & 5007  & 7638 $\pm$ \f 2   &  17 $\pm$ \f 3 & 6.4e-19  &  8    &  17 $\pm$ \f \f 5 & 	             \\ \hline
 5C7.194     &  1.738 &  Ly $\alpha$ / NV  & 1216  & 3352 $\pm$   14   &  92 $\pm$ 30 & 8.9e-17  &  32   & 183 $\pm$ \f 37 &   abs		      \\
7C B081914.4+254809&  $\pm$ &  Si IV / O IV& 1400  & 3841 $\pm$ \f 6   &  97 $\pm$ 18 & 1.0e-17  &  7    &  37 $\pm$ \f 11 &            	      \\
             &   .005 &  C IV              & 1549  & 4242 $\pm$ \f 6   &  85 $\pm$ 16 & 1.5e-17  &  19   &  66 $\pm$ \f 12 &           abs           \\
             &        &  He II             & 1640  & 4464 $\pm$ \f 8   &  99 $\pm$ 19 & 3.3e-18  &  5    &  16 $\pm$ \f \f 7 & 		              \\
             &        &  C III]            & 1909  & 5217 $\pm$ \f 8   & 100 $\pm$ 16 & 1.0e-17  &  16   &  67 $\pm$ \f 13 & 	         	      \\
             &        &  [Ne IV]           & 2424  & 6637 $\pm$ \f 5   &  35 $\pm$ \f 5 & 5.9e-19  &  6    &   5 $\pm$ \f \f 2 & 		              \\
             &        &  Mg II             & 2799  & 7671 $\pm$ \f 6   &  52 $\pm$ 10 & 4.0e-18  &  25   &  42 $\pm$ \f 10 &   atm abs       \\ \hline
\hline         										  
\end{tabular}											  
{\caption[Table of observations]{\label{tab:emli2} (cont)					  
}}
\end{center}              
\normalsize
\end{table*}

\addtocounter{table}{-1}

\begin{table*}
\footnotesize
\begin{center}
\begin{tabular}{lrlccrrrrl}
\hline\hline
\mc{1}{c}{name} &\mc{1}{c}{$z$} &\mc{1}{l}{line} &\mc{1}{c}{$\lambda_{\mathrm rest}$} &\mc{1}{c}{$\lambda_{\mathrm obs}$} &\mc{1}{c}{FWHM}           &\mc{1}{c}{flux}  &\mc{1}{c}{snr} &\mc{1}{c}{W$_{\lambda}$}   &\mc{1}{l}{notes} \\  
\mc{1}{c}{} &\mc{1}{c}{ } &\mc{1}{c}{ }    &\mc{1}{c}{(\AA)}  &\mc{1}{c}{(\AA)} &\mc{1}{c}{(\AA)}  &\mc{1}{c}{(W m$^{-2}$)} &\mc{1}{c}{$\sigma$} &\mc{1}{c}{(\AA)} &\mc{1}{c}{ }\\
\hline\hline
 5C7.195     &  2.034 &  Ly $\alpha$       & 1216  & 3689 $\pm$ \f 4   &  34 $\pm$ \f 4 & 3.9e-18  &  23   & 527 $\pm$ 120 &           	      \\
7C B081920.1+281538  &  $\pm$ &  N V       & 1240  & 3764 $\pm$ \f 4   &  27 $\pm$ \f 5 & 1.4e-18  &  11   & 183 $\pm$ \f 70 &          	      \\
             &   .002 &  Si IV / O IV      & 1400  & 4241 $\pm$   15   &  40 $\pm$   20 & 3.3e-19  &  5    &  41 $\pm$ \f 25 &   cr in line         \\
             &        &  C IV              & 1549  & 4702 $\pm$ \f 2   &  34 $\pm$ \f 3 & 8.5e-19  &  17   &  95 $\pm$ \f 40 &   abs 		      \\
             &        &  He II             & 1640  & 4966 $\pm$ \f 2   &  21 $\pm$ \f 4 & 3.0e-19  &  10   &  33 $\pm$ \f 16 &         		      \\
             &        &  C III]            & 1909  & 5772 $\pm$ \f 5   &  41 $\pm$ \f 7 & 9.2e-19  &  13   & 167 $\pm$ \f 46 &   abs                \\ \hline
5C7.230      &  1.242 &  C IV              & 1549  & 3468 $\pm$ \f 6   &  51 $\pm$   40 & 1.2e-17  &  15   & 159 $\pm$ \f 80 &   in noise	      \\
7C B082134.3+244829  &  $\pm$ &  He II     & 1640  & 3674 $\pm$ \f 5   &  52 $\pm$   40 & 1.7e-18  &  2.8  &  26 $\pm$ \f \f 6 &   in noise	      \\
             &   .002 &	 C III]            & 1909  & 4274 $\pm$ \f 8   &  69 $\pm$   17 & 2.0e-18  &  7    &  39 $\pm$ \f 16 & 			      \\
             &        &  C II]             & 2326  & 5222 $\pm$ \f 2   &  25 $\pm$ \f 5 & 3.1e-19  &  2.6  &   7 $\pm$ \f \f 3 & 		              \\
             &        &  Mg II             & 2799  & 6276 $\pm$ \f 4   &  52 $\pm$   30 & 1.5e-18  &  7    &  40 $\pm$ \f \f 9 &   abs \& cr in line   \\ \hline
7C0825+2930  &  2.315 &  Ly $\alpha$       & 1216  & 4053 $\pm$ \f 8   & 127 $\pm$   18 & 9.0e-18  &  30   & 654 $\pm$ 140 &   abs		      \\
7C B082505.4+293018&  $\pm$ &  Si IV / O IV& 1400  & 4640 $\pm$ \f 6   &  80 $\pm$   22 & 9.1e-19  &  5    &  67 $\pm$ \f 18 & 	              \\
             &   .001 &  C IV              & 1549  & 5137 $\pm$ \f 4   & 144 $\pm$   24 & 5.7e-18  &  43   & 440 $\pm$ 110 & 		              \\
             &        &  C III]            & 1909  & 6328 $\pm$ \f 6   & 186 $\pm$   32 & 2.9e-18  &  22   & 287 $\pm$ \f 70 &   abs		      \\
\hline\hline         
\end{tabular}
\end{center}              
{\caption[Table of observations]{\label{tab:emli3} (cont.) Table
summarising the emission lines present in our optical spectra of the
7C quasars and BLRGs. Emission lines labelled `abs' have absorption in
them. Emission lines labelled `atm abs' and `cr in line' suffer from
atmospheric absorption and cosmic ray hits, respectively. The line
fluxes from the non-spectrophotometric observations in Table 2 have
been scaled by up to 2.5 mag. to account for absorption by cloud.}}
\normalsize
\end{table*}

\clearpage

\addtocounter{figure}{-1}

\begin{figure*}
\epsfxsize=0.9\textwidth
\hspace{0.45cm}
\begin{centering}
\epsfbox{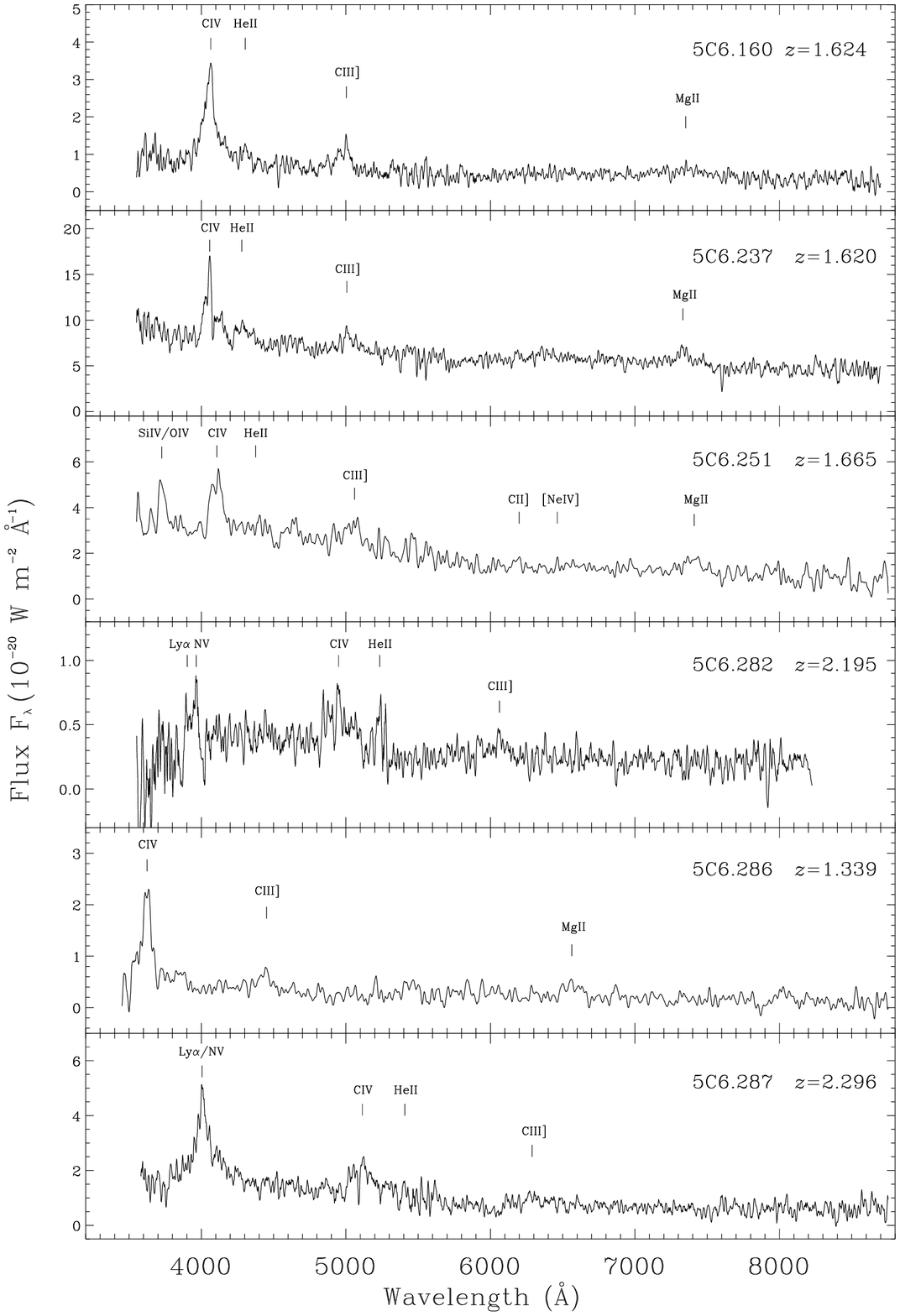} 
\end{centering}
\vspace{1.0cm}
{\caption[junk]{\label{fig:findnsd} (cont)
}}
\end{figure*}					

\addtocounter{figure}{-1}

\begin{figure*}
\epsfxsize=0.9\textwidth
\hspace{0.25cm}
\begin{centering}
\epsfbox{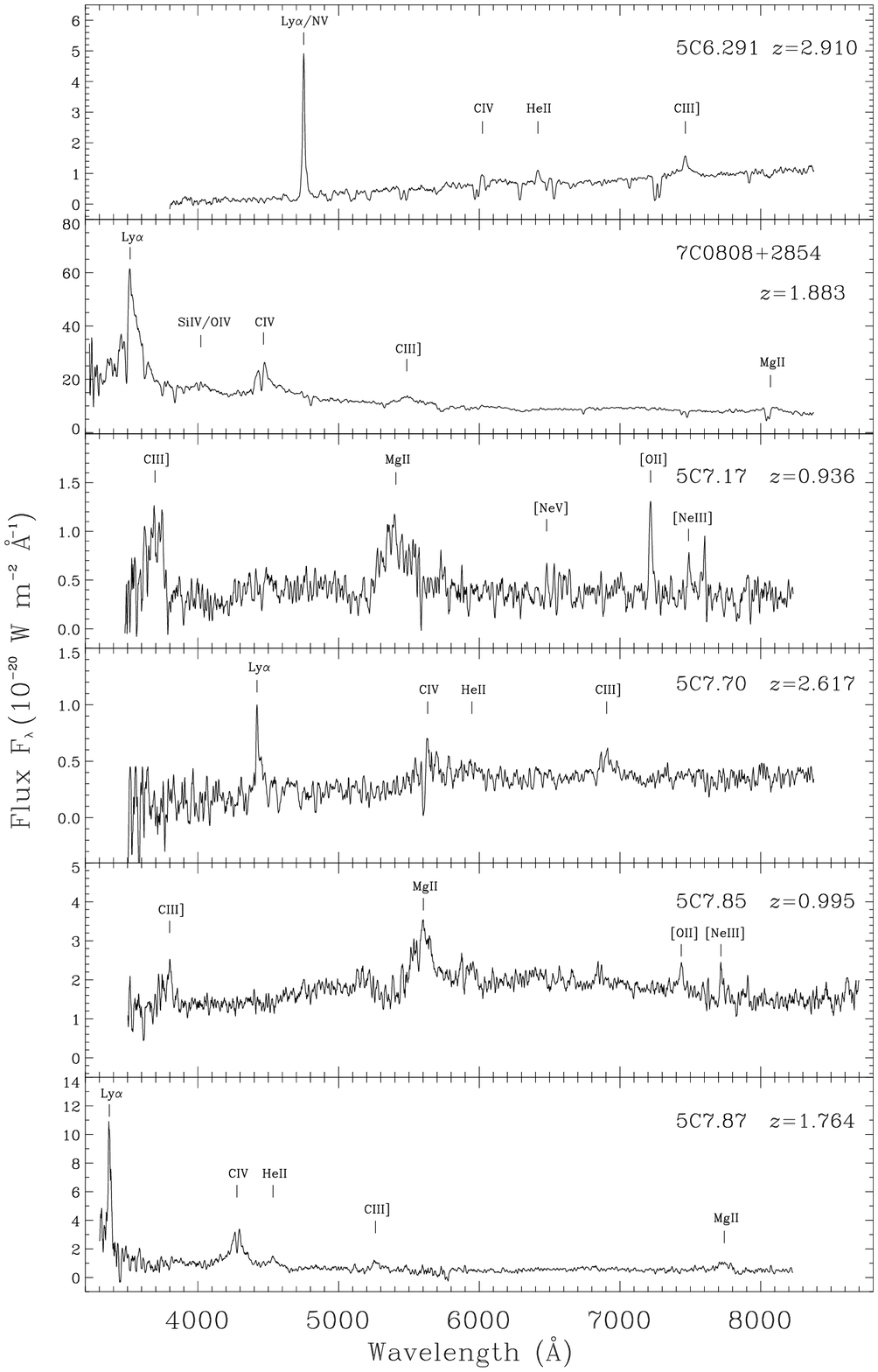} 
\end{centering}
\vspace{1.0cm}
{\caption[junk]{\label{fig:findn} (cont)
}}
\end{figure*}				

\addtocounter{figure}{-1}

\begin{figure*}
\epsfxsize=0.9\textwidth
\hspace{0.45cm}
\begin{centering}
\epsfbox{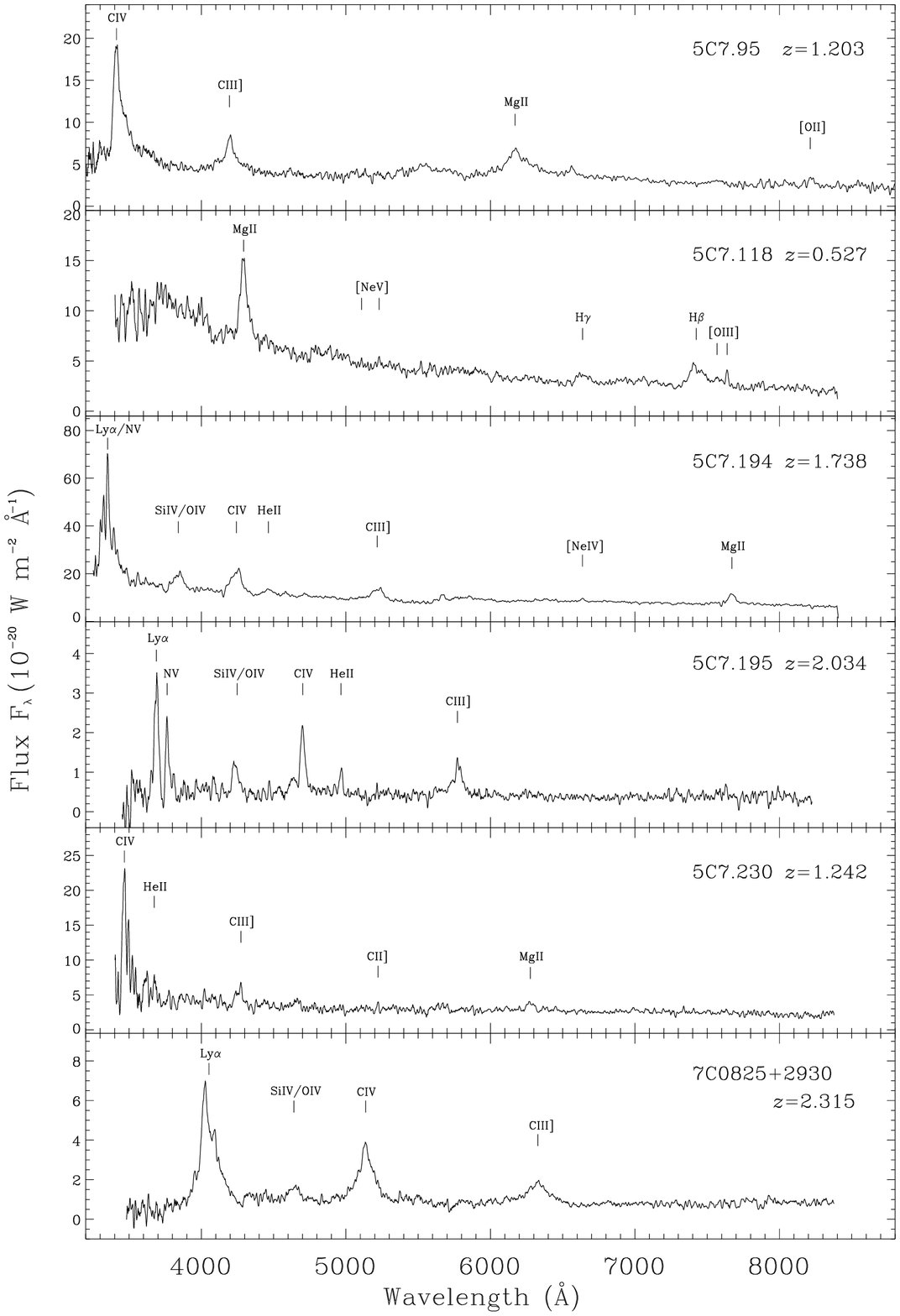} 
\end{centering}
\vspace{1.0cm} {\caption[junk]{\label{fig:findnod} (cont) Optical
spectra of quasars and broad-line radio galaxies in the 7C sample.}}
\end{figure*}

\clearpage
\section{3CRR quasars}
Table B1 lists the 3CRR quasar sample used in this paper which was
combined with the 7C sample to give good coverage of the radio
luminosity--redshift plane. Only two quasars (3C345 and 3C454.3) are
excluded on the basis of having Doppler boosted cores/jets which raise
their total low-frequency flux above the sample limit. One quasar
(3C286) has a rest-frame 1 GHz spectral index $\alpha_{\mathrm rad} <
0.5$, but has enough extended flux to remain in the sample. Therefore
it is called a core-jets quasar (CJS; see Section 2). 

The full list of objects we classify as broad-line radio galaxies in
the 3CRR sample is 3C22 (RLSE), 3C33.1 (LJWU), 3C41 (SRL), 3C61.1
(EH), 3C67 (LJWU), 3C219 (LJWU), 3C268.3 (LJWU), 3C303 (EH), 3C318
(SS), 3C381 (GO), 3C382 (EH), 3C390.3 (EH) \& 3C455 (SDMA). Key to
broad-emission line references: EH - Eracleous \& Halpern (1994); GO -
Grandi \& Osterbrock (1978); LJWU - Laing et al. (1994); RLSE -
Rawlings et al. (1995); SDMA - Spinrad et al. (1985); SRL - Simpson,
Rawlings \& Lacy (in prep.); SS - Spinrad \& Smith (1976)

\begin{table*}
\footnotesize
\begin{center}
\begin{tabular}{lcccclc}
\hline\hline \mc{1}{l}{Name}
&\mc{1}{c}{$S_{178}$}&\mc{1}{c}{$\alpha_{\mathrm rad}$}&\mc{1}{
c}{$z$}&\mc{1}{c}{$\log_{10}(L_{151})$}
&\mc{1}{l}{$~~V$}&\mc{1}{c}{$M_{B}$}\\ \mc{1}{l}{ } &\mc{1}{c}{(Jy)}
&\mc{1}{c}{ } &\mc{1}{c}{ } &\mc{1}{c}{(W Hz$^{-1}$
sr$^{-1}$)}&\mc{1}{c}{} &\mc{1}{c}{}\\ \hline\hline

3C9      &   19.402  & 1.08   &   2.012  &  28.68   &  18.21  &  $-26.60$ \\
3C14     &   11.336  & 0.97   &   1.469  &  28.10   &  20.00  &  $-24.42$ \\
3C43     &   12.644  & 0.76   &   1.457  &  28.09   &  20.00  &  $-23.70$ \\
3C47     &   28.776  & 0.98   &   0.425  &  27.39   &  18.10  &  $-23.57$ \\
3C48     &   59.950  & 0.71   &   0.367  &  27.46   &  16.20  &  $-25.27$ \\
3C68.1   &   13.952  & 0.97   &   1.238  &  28.03   &  19.60  &  $-25.84$ \\
3C109    &   23.544  & 0.79   &   0.305  &  26.96   &  17.88  &  $-23.14$ \\
3C138    &   24.198  & 0.56   &   0.759  &  27.65   &  17.90  &  $-25.00$ \\
3C147    &   65.945  & 0.60   &   0.545  &  27.80   &  16.90  &  $-25.43$ \\
3C175    &   19.184  & 1.01   &   0.768  &  27.77   &  16.60  &  $-26.66$ \\
3C181    &   15.805  & 0.93   &   1.382  &  28.16   &  18.92  &  $-25.75$ \\
3C186    &   15.369  & 1.19   &   1.063  &  27.94   &  17.60  &  $-26.47$ \\
3C190    &   16.350  & 0.88   &   1.197  &  28.06   &  20.00  &  $-23.98$ \\
3C191    &   14.170  & 1.00   &   1.952  &  28.52   &  18.65  &  $-25.86$ \\
3C196    &   74.338  & 0.88   &   0.871  &  28.38   &  17.60  &  $-26.24$ \\
3C204    &   11.445  & 1.07   &   1.112  &  27.94   &  18.21  &  $-26.20$ \\
3C205    &   13.734  & 1.00   &   1.534  &  28.23   &  17.62  &  $-27.46$ \\
3C207    &   14.824  & 0.72   &   0.684  &  27.48   &  18.15  &  $-24.80$ \\
3C208    &   18.312  & 1.07   &   1.109  &  28.06   &  17.42  &  $-26.49$ \\
3C212    &   16.459  & 0.79   &   1.049  &  27.92   &  19.06  &  $-24.73$ \\
3C215    &   12.426  & 1.02   &   0.411  &  26.98   &  18.27  &  $-23.35$ \\
3C216    &   22.018  & 0.67   &   0.668  &  27.65   &  18.48  &  $-24.27$ \\
3C245    &   15.696  & 0.73   &   1.029  &  27.88   &  17.25  &  $-26.73$ \\
3C249.1  &   11.663  & 0.88   &   0.311  &  26.69   &  15.72  &  $-25.32$ \\
3C254    &   21.691  & 1.05   &   0.734  &  27.74   &  17.98  &  $-24.75$ \\
3C263    &   16.568  & 0.87   &   0.656  &  27.50   &  16.32  &  $-26.22$ \\
3C268.4  &   11.227  & 0.90   &   1.400  &  28.01   &  18.42  &  $-26.72$ \\
3C270.1  &   14.824  & 0.80   &   1.519  &  28.23   &  18.61  &  $-25.54$ \\
3C275.1  &   19.947  & 0.82   &   0.557  &  27.52   &  19.00  &  $-23.26$ \\
3C286    &   27.250  & 0.37   &   0.849  &  27.58   &  17.25  &  $-25.94$ \\
3C287    &   17.767  & 0.52   &   1.055  &  27.79   &  17.67  &  $-26.78$ \\
3C309.1  &   24.743  & 0.61   &   0.904  &  27.85   &  16.78  &  $-26.90$ \\
3C334    &   11.881  & 0.90   &   0.555  &  27.25   &  16.41  &  $-25.78$ \\
3C336    &   12.535  & 0.92   &   0.927  &  27.72   &  17.47  &  $-26.24$ \\
3C343    &   13.516  & 0.68   &   0.988  &  27.60   &  20.61  &  $-23.27$ \\
3C351    &   14.933  & 0.78   &   0.371  &  26.91   &  15.28  &  $-26.11$ \\
4C16.49  &   11.445  & 1.00   &   1.296  &  28.05   &  18.50  &  $-25.62$ \\
3C380    &   64.746  & 0.72   &   0.691  &  28.10   &  16.81  &  $-25.91$ \\
3C432    &   11.990  & 1.07   &   1.805  &  28.36   &  17.96  &  $-26.61$ \\
3C454    &   12.644  & 0.79   &   1.757  &  28.33   &  18.47  &  $-25.86$ \\
								      
\hline\hline         						      
\end{tabular}		
\end{center}              
{\caption[Table of observations]{\label{tab:3Ctab} Data for the 3CRR
quasar sample. The 178 MHz fluxes are those listed in LRL multiplied
by 1.09, to correct for an error in the KPW flux scale at 178 MHz
(Roger et al. 1973). Spectral indices have been evaluated at 1 GHz
rest-frame. Rest-frame 151 MHz luminosities are calculated from
corrected 178 MHz fluxes and spectral fitting as described in Blundell
et al. (1998a).  All $V$ magnitudes are taken from LRL and have not been
corrected for galactic reddening. Absolute magnitudes were calculated
as described in Section 5.1. L$_{151}$ and $M_{B}$ assume
$\Omega_{M}=1$, $\Omega_ {\Lambda}=0$.}}  \normalsize
\end{table*}

\end{document}